\pdfoutput=1
\documentclass[%
 reprint,
 amsmath,amssymb,
 aps,
 prx,
]{revtex4-2}

\usepackage{graphicx}% Include figure files
\usepackage{dcolumn}% Align table columns on decimal point
\usepackage{bm}% bold math
\usepackage{xcolor}% Colored text for highlighting
\begin{document}

\preprint{}
\title{Measurements as a roadblock to near-term practical quantum advantage in chemistry: resource analysis}

\author{J\'er\^ome F. Gonthier$^{1}$}
\email{jerome@zapatacomputing.com}
\author{Maxwell D. Radin$^1$}
\author{Corneliu Buda$^2$}
\author{Eric J. Doskocil$^2$}
\author{Clena M. Abuan$^3$}
\author{Jhonathan Romero$^1$}
\affiliation{$^{1}$Zapata Computing, Inc., 100 Federal St., Boston, MA 02110, USA}
\affiliation{$^{2}$BP Innovation and Engineering,
150 West Warrenville Road, Naperville, IL 60563, USA}
\affiliation{$^3$BP Innovation and Engineering,
501 Westlake Park Blvd, Houston, TX 77079, USA}

\date{\today}% It is always \today, today,

\begin{abstract}
Recent advances in quantum computing devices have brought attention to hybrid quantum-classical algorithms like the Variational Quantum Eigensolver (VQE) as a potential route to practical quantum advantage in chemistry.
However, it is not yet clear whether such algorithms, even in the absence of device error, could actually achieve quantum advantage for systems of practical interest.
We have performed an exhaustive analysis to estimate the number of qubits and number of measurements required to compute the combustion energies of small organic molecules and related systems to within chemical accuracy of experimental values using VQE. We consider several key modern improvements to VQE, including low-rank factorizations of the Hamiltonian. Our results indicate that although these techniques are useful, they will not be sufficient to achieve practical quantum computational advantage for our molecular set, or for similar molecules. 
This suggests that novel approaches to operator estimation leveraging quantum coherence, such as Enhanced Likelihood Functions, may be required.
\end{abstract}

\maketitle

\section{Introduction}
\label{sec:Introduction}

In the last decade, quantum computers have evolved from laboratory prototypes of a few qubits to machines with tens of qubits that are commercially available for researchers and businesses to use \cite{castelvecchi2017ibm,mohseni2017commercialize}. In 2019, Google announced the realization of the quantum supremacy milestone: their 53-qubit chip accomplished a specific task that would be extremely difficult to simulate with a classical supercomputer \cite{Arute2019}, though recent work improved considerably on initial classical runtime estimates \cite{supremacy_simulation}. This task was specifically designed to be well suited to the quantum processor and challenging for classical computers, and does not solve a practical problem. The next milestone, and arguably the most pressing one \cite{Grumbling2019}, is finding a \emph{practical quantum advantage} with noisy intermediate-scale quantum (NISQ) devices \cite{Preskill2018}, that is, running an algorithm on a NISQ device that provides an improved solution for a commercially relevant task. This improvement can manifest in different ways, either as a reduction in the time to solution or an increase in the quality of the solution. Accomplishing this goal requires first a steady improvement in the quality of quantum computing hardware. Fortunately, we are witnessing a rapid growth in the number of qubits and fidelity of these machines as indicated by the recent trends in metrics such as quantum volume \cite{Cross2019}: in the past couple of years this went from 32 in January 2020 \footnote{IBM blog January 2020: https://www.ibm.com/blogs/research/2020/01/quantum-volume-32/}, to 128 in September 2020 \footnote{Honeywell press release September 2020: https://www.honeywell.com/en-us/newsroom/news/2020/09/achieving-quantum-volume-128-on-the-honeywell-quantum-computer}, then 2048 in December 2021 \footnote{Quantinuum press release December 2021: https://www.quantinuum.com/pressrelease/demonstrating-benefits-of-quantum-upgradable-design-strategy-system-model-h1-2-first-to-prove-2-048-quantum-volume} and 4096 in April 2022 \footnote{Quantinuum press release April 2022: https://www.quantinuum.com/pressrelease/quantinuum-announces-quantum-volume-4096-achievement}. The second requirement towards quantum advantage is identifying commercially relevant tasks for which a near-term quantum algorithm can provide a measurable improvement compared to classical alternatives.

Quantum chemistry has been identified as a likely candidate \cite{aspuru2005simulated,cao2019quantum,McArdle2019rev} for quantum advantage for multiple reasons. First, electronic structure calculations are used extensively in the development of many technologies, for example in the chemicals industry \cite{Deglmann2015}, drug development \cite{Heifetz2020}, and battery materials research \cite{VanderVen2020}.
Second, electronic structure calculations rely on the Schr{\"o}dinger equation, for which a general exact solution has exponential cost on a classical computer with all known classical methods. Third, quantum computers can store exponentially scaling representations of the wavefunction on a linear number of qubits and also provide means to implement Hamiltonian evolution efficiently, giving rise to quantum algorithms to estimate ground state energies of some molecular Hamiltonians using polynomially scaling resources.

Quantum approaches to electronic structure calculations can be divided into two categories: 1) algorithms based on the Quantum Phase Estimation subroutine and related techniques and 2) quantum heuristic algorithms \cite{cao2019quantum}, such as the variational quantum eigensolver (VQE) \cite{Peruzzo2014} and related methods based on different versions of the time-dependent variational principle \cite{McArdle2019rev}. Under certain assumptions, approaches in the first category can provide an advantage in computational scaling compared to exact classical algorithms, however they
require a fault-tolerant implementation \cite{aspuru2005simulated,reactions_on_QC,Burg2020,gate_count_QPE,QPE_trotterization,lee2020efficient}, and therefore are not applicable in the near-term.
In contrast, quantum heuristics such as VQE can be implemented on NISQ devices thanks to the flexibility in their construction, but do not provide any proven asymptotic advantage compared to classical algorithms. Demonstrating a quantum advantage in this context requires a comparison of the computational cost and performance of the quantum heuristic against the state-of-the-art classical approximations for specific problem instances. Performance metrics might include aspects such as total runtime and accuracy of the solution. In addition, this comparison must go hand in hand with an optimization of the algorithmic choices of the heuristic to maximize performance. From this perspective, a necessary step towards achieving quantum advantage in the near term is establishing protocols for evaluating quantum computational resources for specific sets of problem instances and target accuracy of the solution, a procedure we will refer to as Resource and Performance Assessment (RPA).

While many studies have estimated resource requirements for quantum chemistry using fault-tolerant algorithms such as Quantum Phase Estimation \cite{reactions_on_QC,Burg2020,gate_count_QPE,QPE_trotterization,industrial_Q_advantage,lee2020efficient},
only a small number have assessed the resource requirements for NISQ approaches.
McClean \textit{et al.} have analyzed the asymptotic measurement requirements of VQE \cite{meas_scaling_bounds} while K{\"u}hn \textit{et al.} numerically examined qubit requirements and required circuit depth for UCC-derived Ans\"atze \cite{Kuehn2019}. Numerical studies have explored the VQE measurement costs of diatomic molecules and hydrogen chains or rings when applying fermionic marginal constraints and a low-rank factorization of the Hamiltonian \cite{marginal_constraints,Huggins_tens_fact}.
Elfving et al. estimated the number of spin-orbitals required for industrially-relevant calculations and concluded that the required VQE execution time was prohibitively large, although the method used to estimate the execution time was not specified \cite{industrial_Q_advantage}.

While these studies provide valuable insight into the performance of VQE and its variants, several questions about the feasibility of these techniques for systems of practical interest remain unanswered.
For example, although VQE measurement requirements have been assessed for small basis sets, such analysis has not been carried out for the basis sets required to achieve a useful accuracy with respect to the infinite basis set limit.
Another key question is how these measurement requirements compare to the runtime of state-of-the-art classical quantum-chemistry techniques.
Furthermore, previous studies have estimated measurements by employing canonical orbitals, and have not considered Frozen Natural Orbitals (FNO)\cite{FNO_CI_first,reduced_virt_space_FNO,FNO_for_CC}, which are known to significantly reduce both the computational cost of classical wavefunction-based quantum chemistry methods and the requirements on the number of qubits for quantum computing applications \cite{Verma_FNO,LiH_NOs_edu}.

To address these questions, we have performed an RPA to estimate the number of qubits, number of measurements, and total runtime required for calculating combustion energies for small organic molecules to within chemical accuracy (defined as 4.2~kJ/mol or 1.6~mHa \cite{chem_acc_2012}) with a single VQE energy evaluation, leaving the problem of VQE parameter optimization to future work. These estimates consider Frozen Natural Orbitals as well as measurement reduction techniques such as Hamiltonian grouping of commuting \cite{Greedy_grouping_algo,Kandala2017,min_clique_grouping,linear_grouping_algo,flow_network_grouping} or anti-commuting \cite{anticommuting_groups,zhao.PhysRevA.101.062322} terms, the application of fermionic marginal constraints \cite{marginal_constraints}, and low-rank factorization of the Hamiltonian \cite{OF_patent,Huggins_tens_fact}.

Our results indicate that between 120 and 260 qubits are required for chemical accuracy for our benchmark systems. Under optimistic assumptions about the Ansatz requirements and the sampling rate of the device, we show that a single energy evaluation could take several days to weeks, rendering the calculations impractical and inferior to classical methods, in particular when considering the large number of such evaluations required for the optimization loop of VQE. Our results also show that although certain grouping techniques greatly reduce the number of measurements, they are not enough to guarantee practical runtimes in the regime where quantum advantage is expected. This suggests that making VQE practical in the near-term requires the use of new approaches to measurement that leverage quantum coherence to reduce estimation runtimes, such as the  recently proposed Bayesian Likelihood Function techniques \cite{Bulbasaur_1,ELF}.

The rest of the paper is organized as follows: in Section \ref{sec:methods}, we describe our methods for each step of the RPA in details.  In Section \ref{sec:results}, we present the numerical results for our RPA. There, we establish accurate classical quantum chemistry reference values by comparison with experimental reaction energies. We then truncate the active space to establish the minimal number of qubits necessary to preserve chemical accuracy. The last step of our estimation evaluates the number of necessary measurements to reach chemical accuracy on a quantum computer, including measurement reduction techniques. This evaluation is made for increasing active space sizes to establish asymptotic scaling relationships. Finally, we introduce an empirical extrapolation formula to establish runtime and resource requirements for more general systems than the ones specifically studied in this work. In the final Section \ref{sec:conclusions}, we discuss our results, their implications and further research avenues.

\section{Methods}
\label{sec:methods}

In this section we describe our methodology for resource estimation, starting with an outline of the RPA concept as applied to VQE, followed by a detailed description of the methods employed for the estimation of classical and quantum computational resources. All calculations were deployed using Zapata Computing's Orquestra\textsuperscript{\textregistered} workflow management platform.

\subsection{Outline of the resource and performance assessment}

The goal of an RPA is simple: we want to estimate as accurately as possible the resources, such as number of qubits, number of measurements, fidelity, among others, needed to achieve a given quality of solution for a specific choice of quantum algorithm and a set of problem instances. By fixing a target quality in the solution, it is possible to compare the cost with that of state-of-the-art classical approaches, establishing whether a quantum advantage is possible. The process can be divided in five stages:
\begin{enumerate}
    \item Define a set of problem instances and the quantum algorithm to be assessed.
    \item Set a target metric for performance. For example, choose a target quality of solution or time to solution.
    \item Select a classical approach for comparison, ideally the state-of-the-art method for the instances of interest, and estimate the amount of classical resources required to achieve the target performance.
    \item Estimate the amount of quantum computational resources required to achieve the target performance using the quantum algorithm.
    \item Compare the performances and computational cost of the quantum and classical approaches. Establish whether a practical quantum advantage is attainable. 
\end{enumerate}

In this work, the problem under study is the calculation of the combustion energy for a set of small organic molecules. We selected the gold standard for quantum chemistry, Coupled-Cluster with Singles, Doubles and perturbative Triples or CCSD(T) as our classical algorithm, and evaluated the cost of estimating combustion reaction energies to chemical accuracy. This resource estimate includes determining the basis set and number of spin-orbitals needed. The performance of the classical approach provides the reference to be outperformed by the quantum algorithm, setting chemical accuracy as the target metric for quantum advantage. In principle, it is also possible to set experimental data as the reference, provided reliable enthalpic corrections to the electronic energies are available. In our case, we instead ensured that the classical approach was reproducing experimental data, and then used the classical electronic energy results as reference for the quantum algorithms.

With the target accuracy fixed, we proceeded to estimate the number of qubits and total number of measurements needed to achieve such accuracy assuming access to a sufficiently expressive Ansatz and high enough gate fidelity. By incorporating assumptions about the characteristics of the variational circuit and the quantum hardware, we established realistic runtime estimates for achieving chemical accuracy. Crucially, our analysis takes into account the system and size dependence of different resource requirements. For this reason, we focus on techniques that can be scaled up to larger molecules or clusters, which is why considerations of spatial symmetry for example are not included.

While the most accurate RPA would require executing the algorithm, this might be too computationally costly. We can instead take advantage of our knowledge of the algorithm and the problem to investigate empirical scaling of the resources with system size in order to establish relationships that allow extrapolation to larger instances. Some performance metrics, such as the number of measurements in VQE, only depend on properties of the problem instance. More specifically, the number of measurements to reach a target accuracy can be predicted from the Hamiltonian of each molecule and classical estimations for variances (see Section \ref{subsubsec:meas_analysis}). In the rest of this section we describe in details the benchmark data set chosen for our RPA as well as the methodology for estimating classical and quantum computational resources.

\subsection{Benchmark data set}

In this work, we aim to establish resource estimates for computing electronic ground state energies with VQE. We wish to apply our resource estimation procedure to a benchmark set of molecules to facilitate extrapolation to larger systems. Ideally, this set would be of practical relevance, contain small enough molecules to allow chemically accurate classical computations, and correspond to well-established, accurate experimental data. For these reasons, we chose to study combustion reactions for the following small hydrocarbons: methane, methanol, ethane, ethene, ethyne, ethanol, propane, propene, and propyne (depicted in Figure %\ref{fig:molecule_set} 
S1 in \cite{Supp_mat}). For clarity, we explicitly write the general formula for the reaction's stoichiometry:
\begin{equation}
    \text{C}_{x}\text{H}_{y}\text{O}_{z} + \left(x + \frac{y}{4} - \frac{z}{2}\right) \text{O}_{2} \rightleftharpoons x \text{CO}_{2} + \frac{y}{2} \text{H}_{2}\text{O}
\end{equation}

Experimental enthalpies of combustion for the hydrocarbons in our benchmark set can easily be calculated from available enthalpies of formation \cite{CRC_96} (see Table %\ref{tab:experimental_enthalpies}
S1 \cite{Supp_mat}). By combining electronic ground state energies with vibrational, rotational, and translational enthalpic contributions, we can obtain simulated combustion enthalpies that can be compared to the experimental values. Most of our work focuses on getting accurate electronic energies, as harmonic vibrational corrections to enthalpies are obtained from the second derivatives of the electronic energies. Anharmonic effects are expected to be important for larger, flexible molecules but only play a very minor role in our benchmark reactions, as numerically verified in Table %\ref{tab:composite_components}
S4 \cite{Supp_mat}.

Algorithms to compute anharmonic vibrational spectra on quantum computers exist \cite{anharm_vib_QC,vib_wft_QC,annealer_vib_calc,HEA_vibrational_QC}, and have been argued to be better candidates than electronic structure for early quantum advantage \cite{vib_quant_advantage}. This assessment was based on considerations of scaling of the number of terms and their locality in the respective Hamiltonians. In addition, the relative magnitude of the Hamiltonian coefficients also favored vibrational Hamiltonians for the systems considered. We hope that our method for resource estimation provides an accurate picture of the prospects of electronic structure algorithms on quantum computers for concrete examples, thus facilitating the comparison with prospects of quantum algorithms for vibrational structure.

Our chosen set of molecules is dominated by dynamical correlation. As pointed out by Elfving \textit{et al.}, this means that a very large number of orbitals is needed for accurate treatment \cite{industrial_Q_advantage}. Hence, a very large number of qubits would be needed on a quantum computer to rival with quantum chemistry capabilities on classical computers. In that sense, systems dominated by non-dynamical correlations would be better candidates for demonstrations of near-term quantum advantage. However, we believe that most of our extrapolation and resource estimation results are valid for general molecular systems, whether dominated by dynamical or by non-dynamical correlations. In particular, our results regarding the scaling of the number of measurements necessary to reach chemical accuracy with the size of the system should be transferable to most cases.

\subsection{Methodology for resource estimation}

\subsubsection{Classical benchmarks}

The first component of the RPA consists of establishing a reference classical quantum chemistry approach and evaluating the classical resources needed to achieve chemical accuracy for a benchmark set of molecules.
The current gold standard for ground state electronic structure calculations is the Coupled-Cluster with Singles, Doubles and perturbative Triples CCSD(T) method \cite{CCSD_T_method}. For closed-shell molecules, and with sufficiently large basis sets, CCSD(T) can reach chemical accuracy, i.e. an error of 4.2~kJ/mol compared to experimental data \cite{chem_acc_2012,bench_NCI_Hobza}. Unfortunately, its $N^7$ scaling limits its application to small systems. For completeness, we note that approximate CCSD(T) methods were developed \cite{DLPNO_CCSD_method,DEC_CCSD_T,OSV_LCCSD,Kallay_LNO_CCSD_T} that take advantage of the spatial locality of electron correlation, which allowed computation of much larger systems \cite{Kallay_LNO_CCSD_T,DLPNO_CCSD_appl} by reducing the overall asymptotic scaling. However, the accuracy of the local approximations in coupled-cluster methods was recently questioned \cite{large_mol_CC_error,CC_large_ion_error}.

In this work, we assessed the suitability of CCSD(T) for our benchmark set. We also explored which basis set was large enough to reach chemical accuracy for CCSD(T), by performing computations in various aug-cc-pVXZ basis sets (denoted AVXZ) with X = D, T, Q or 5 \cite{aug-cc-pVXZ} denoting increasing angular momentum of the basis functions, and in the def2-TZVPPD basis set \cite{def2-basis-sets,def2_diffuse}. The details of our computational methods and results for this assessment are reported in the Supplemental Material \cite{Supp_mat} for the interested reader.

\subsubsection{Number of qubits}
\label{subsubsec:nqubits_summary}

To estimate the minimal number of qubits that can be used while recovering CCSD(T)/AV5Z results with sufficient accuracy, we explored different choices of orbitals and active spaces. In practice, we evaluated the CCSD(T)/AV5Z energy with and without truncations and we assumed that the resulting errors are reflective of those that would be obtained with an accurate VQE Ansatz.

The simplest choice for orbital truncation is to eliminate Hartree-Fock canonical virtual (i.e. unoccupied) orbitals with the highest energy. However, a better choice is known in the quantum chemistry literature: the Frozen Natural Orbitals (FNO) \cite{FNO_CI_first,reduced_virt_space_FNO,FNO_for_CC} method. FNOs have been successfully applied to reduce the number of qubits needed in quantum chemistry simulations on quantum computers \cite{Verma_FNO,LiH_NOs_edu}, however the corresponding number of measurements has to our knowledge not been estimated. As is usual, we apply a perturbation theory correction to partly compensate for the truncated energy, for both canonical and Frozen Natural Orbitals. Further technical details are reported in the Supplemental Material \cite{Supp_mat}.

We also note that there exist other methods to reduce the number of qubits necessary to encode a problem: improved basis sets \cite{MRA_VQE}, exploitation of symmetries \cite{bravyi2017tapering,Steudtner2018}, or partitioning methods \cite{entanglement_forging, DMET_IonQ}. We leave their study to future work.

\subsubsection{Measurement analysis}
\label{subsubsec:meas_analysis}

\paragraph{General considerations} To obtain the energy in the Variational Quantum Eigensolver (VQE) algorithm, it is necessary to estimate the expectation value of the Hamiltonian by performing many measurements and averaging their results. The total number of necessary measurements $M$ can be estimated as follows:
\begin{equation}
    M = \frac{K}{\epsilon^2},
\end{equation}
where $\epsilon$ is the desired precision on the estimation and $K$ is a proportionality constant that depends on the Hamiltonian, the state being measured and the measurement strategy employed for the estimation \cite{McClean2016,meas_scaling_bounds}, as described below. 
Note that in quantum chemistry, an accuracy of 1.6~mHa with respect to the exact ground state in the infinite basis set limit is typically desired, which means that the uncertainty due to sampling error $\epsilon$ must be less than this amount.

After transformation to the qubit representation, the molecular Hamiltonian takes the form:
\begin{equation}
    \hat{H} = \sum_{i} h_{i} \hat{P}_{i},
\end{equation}
where $\hat{P}_{i}$ is a product of Pauli operators acting on one or more qubits and $h_{i}$ is the associated coefficient, obtained from the one- and two-electron integrals calculated with Psi4 \cite{Psi4}.

While the simplest approach to estimating the expectation value of $\hat{H}$ would be to measure each $\hat{P}_i$ independently, it is possible to measure two operators $\hat{P}_i$ and $\hat{P}_j$ at the same time if they commute, thereby reducing the total number of measurements needed. In general, measuring two commuting operators $\hat{P}_i$ and $\hat{P}_j$ implies multi-qubit measurements \cite{min_clique_grouping} or appending a unitary transformation to the circuit \cite{full_commuting}. Hence, we first consider grouping methods that do not increase the depth of the circuit and rely only on single-qubit measurements, i.e. Qubit-Wise Commuting (QWC) groups \cite{Kandala2017,min_clique_grouping}. QWC implies that for both $\hat{P}_i$ and $\hat{P}_j$, the Pauli operators acting on the same qubit individually commute. Finding the optimal QWC grouping is equivalent to solving the Minimum Clique Cover graph problem and is NP-hard in the general case \cite{min_clique_grouping}. Here we use a heuristic greedy algorithm that goes through all operators and adds each one to the first group with which it is Qubit-Wise Commuting \cite{Greedy_grouping_algo,smith2016practical,OpenFermion}. In addition, we sort the list of operators according to their coefficients $h_i$, so operators with the largest coefficients are grouped first \cite{sorted_insertion}.

We also consider the basis rotation approach to Hamiltonian decomposition \cite{OF_patent,Huggins_tens_fact, OpenFermion}.
In the variant applied in this work, the Hamiltonian terms that only contain $\hat{Z}$ operators are measured in the usual way, while an eigenvalue decomposition is used to obtain a low-rank factorization of the remaining two-body terms.
The expectation values of this low-rank factorization and the remaining one-body terms can be obtained by applying a linear-depth basis rotation circuit after the Ansatz.

Other methods to obtain the expectation value of the Hamiltonian at reduced cost exist \cite{linear_grouping_algo,flow_network_grouping,overlap_grouping, comput_basis_sampling}, for example it is possible to measure groups of fully commuting Pauli terms by appending the appropriate circuit to the Ansatz \cite{full_commuting}, or to use locally-biased \cite{LBCS_method} or derandomized \cite{derandomized_shadows} classical shadows as an alternative to QWC grouping. In the current work, we only assessed the performance of one additional method based on grouping mutually anticommuting Pauli terms \cite{zhao.PhysRevA.101.062322,anticommuting_groups}. Our results show that anticommuting grouping is less performant than QWC grouping (see Figure %\ref{fig:UD_grouping}
S12 \cite{Supp_mat}), therefore these results are not included in the main text. Benchmarking of additional methods like those mentioned above will be the object of future work. A particularly promising approach makes use of overlapping Pauli grouping \cite{overlap_grouping}.

\paragraph{Measurement estimation} The grouped Hamiltonian can be rewritten
\begin{equation}
    \hat{H} = \sum_{C}\sum_{\alpha \in C} h_{\alpha} \hat{P}_{\alpha},
\end{equation}
where $C$ indexes groups and $\alpha$ labels terms in a group. Applying the Lagrangian approach of Rubin \textit{et al.} \cite{marginal_constraints} shows that the optimal allocation of measurements to groups gives the following expression for the proportionality constant $K$:
\begin{equation}
\label{eq:K}
    K = \left( \sum_{C} \sqrt{\sum_{\alpha, \beta \in C} h_{\alpha} h_{\beta} \text{Covar}\left( \hat{P}_{\alpha}, \hat{P}_{\beta} \right) } \right)^2
\end{equation}
This measurement allocation scheme assumes that the covariances between all operators $\hat{P}_{\alpha}$ and $\hat{P}_{\beta}$ are already known,
including the variances $\text{Var}\left(\hat{P}_{\alpha}\right) = \text{Covar}\left( \hat{P}_{\alpha}, \hat{P}_{\alpha} \right)$.
In general, one does not know the values of these covariances and must estimate them.

Because $K$ depends on the variances and covariances of the operators $\hat{P}_i$,  $K$ depends on the quantum state being measured, i.e. it changes through the VQE optimization. To avoid these complications, and allow estimation of $K$ for up to 80 qubits, we employ some simplifying approximations. For the variances, we consider two approximations. In the first, we assume variances to be 1.0, the upper bound. In the second, we estimate variances from Configuration Interaction Singles and Doubles (CISD) density matrices computed with Psi4 \cite{Psi4_CI}. The variances can be approximated from the expectation value $\left\langle \hat{P} \right\rangle$ of each operator since $\hat{P}^2 = 1$:
\begin{equation}
    \text{Var}\left( \hat{P} \right) = \left\langle \hat{P}^2 \right\rangle -  \left\langle \hat{P} \right\rangle^2 = 1 - \left\langle \hat{P} \right\rangle^2
\end{equation}
where we approximate the exact expectation value by its CISD counterpart.

We assume all of the covariances between different terms are zero. This does not correspond to a worst or best case scenario, but approximates the effect of a random distribution of covariances within bounds given by the Cauchy-Schwarz inequality:
\begin{equation}
\label{eq:Cauchy}
    \left| \sqrt{\text{Var} \left( \hat{P}_{\alpha} \right) \text{Var} \left( \hat{P}_{\beta} \right)} \right| \ge \text{Covar}\left( \hat{P}_{\alpha}, \hat{P}_{\beta} \right)
\end{equation}
Indeed, covariances can be positive or negative, but their magnitude is bounded by Equation (\ref{eq:Cauchy}). Hence a random distribution is expected to have an average around zero. That is exactly the case when the quantum state being measured is a Haar random distribution \cite{linear_grouping_algo}.
In practice, we observed that this approximation resulted in $K$ being overestimated by a factor of ${\sim}2$ relative to estimates using covariances obtained from circuit simulations of optimized Ans\"atze.
Note that when the upper bound is used for variances and covariances are set to zero, the estimated value of $K$ is entirely determined by the coefficients of the Hamiltonian.

A precise assessment of the number of measurements required and its scaling with increasing system size for both QWC and basis rotation grouping is fundamental in predicting the runtime of energy estimation on NISQ devices. For this purpose, we computed $K$ for all the molecules in our benchmark set with the exception of O$_2$ for technical reasons. For each molecule, we generated the Hamiltonians from one- and two-electron integrals obtained from Psi4 for different active space sizes, where we always used an integer number of qubits per active electron to facilitate extrapolation. $K$ was computed for all these Hamiltonians with up to 80 qubits, and a power fit of $K$ as a function of the number of qubits was used to extrapolate the number of measurements necessary for the 100 to 200 qubits region. In addition, we performed these estimations for both the upper bound approximation to the variances and variances computed from CISD. Finally, for each case we also computed the Hamiltonian coefficients based on canonical orbitals (as is usual in most VQE publications) and based on FNOs (consistent with our active space size estimations), relying on the aug-cc-pVDZ (denoted AVDZ) Dunning basis set in all cases. This results in a total of eight $K$ estimations for each molecule and active space, giving us unprecedented insight into the relative performance of the variants examined.

\paragraph{Variance reduction}
Grouping the Hamiltonian terms is not the only possibility to reduce the total number of measurements needed. The Hamiltonian can also be transformed so that its overall variance is reduced. Here, we explored the Reduced Density Matrix Constraints (RDMC) method \cite{marginal_constraints} proposed by Rubin \textit{et al.} In brief, this method adds operators to the Hamiltonian that sum to zero and optimizes their coefficients to reduce the total variance. We implemented this method directly in the qubit picture, which was suggested by Rubin \textit{et al.} to have better performance than the original implementation in the fermionic picture. We present a comparison of both implementations in Figure %\ref{fig:RDMC_fermion_qubit} 
S11 \cite{Supp_mat} for the interested reader. We apply RDMC to a small set of molecules, with up to 20 qubits included in the active space, and examine the reduction obtained in $K$ for the case of no grouping, for QWC grouping and for the basis rotation grouping. The Hamiltonians examined were computed with FNOs based on the aug-cc-pVTZ (denoted AVTZ) Dunning basis set and variances were estimated from CISD density matrices.

In conclusion, our resource and performance assessment includes a benchmark of classical methods, which can then be used as a reference to estimate the number of qubits needed to reach chemical accuracy in the general case. We establish empirical scaling relations for the number of measurements using state-of-the-art grouping and measurement reduction techniques, various approximations for the variances involved and two different molecular orbital bases. These scaling relations and their prefactors allow us to estimate the number of measurements needed to reach chemical accuracy when the qubit active space for the molecules in our benchmark reaches 100 to 200 qubits. These relations are also useful as a general guide for the scaling of QWC and basis rotation methods, and for the performance of RDMC.

\section{Results}
\label{sec:results}

\subsection{Benchmarking classical chemistry methods}
\label{subsec:classical_res}

The main purpose of this section is to establish whether classical quantum chemistry methods can reach chemical accuracy for combustion reaction of small, closed-shell hydrocarbons, and to quantify how much effort is necessary to reach chemical accuracy. The results of our assessment of classical resources will establish a reference for the next step of our resource evaluation which is concerned with the number of qubits required for chemical accuracy. 

The first step in this assessment is to check whether affordable computational methods yield accurate enough results, in which case it would be unnecessary to consider more computationally expensive methods. Initial tests showed that current Density Functional Theory functionals and perturbation theory were not accurate enough for our purpose, while the inclusion of harmonic enthalpic contributions is necessary for comparison with experiments. For details, we refer the reader to the Supplemental Material \cite{Supp_mat}. Here, we directly investigate the "gold standard" of quantum chemistry, CCSD(T) complemented by the harmonic enthalpic contributions. In particular, we are looking to select an appropriate basis set for this method.
\begin{figure}
\center
\includegraphics[width=8.6cm]{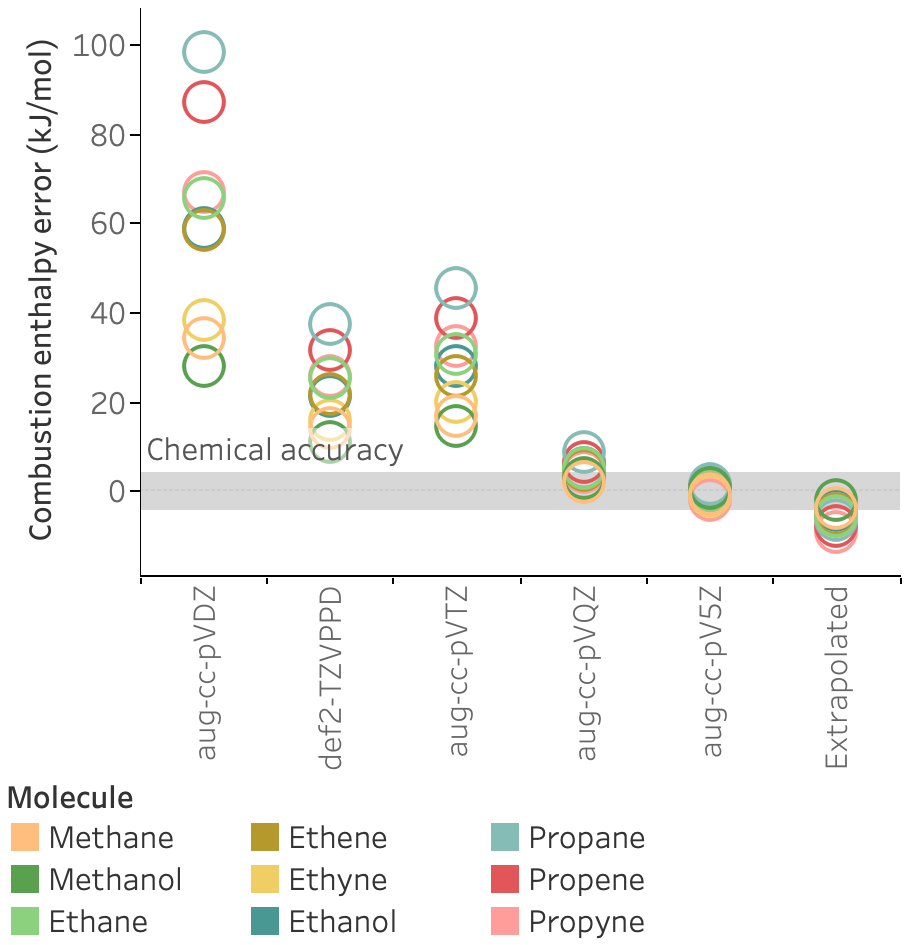}
\caption{CCSD(T) combustion enthalpy errors in kJ/mol in various basis sets and with AVQZ/AV5Z extrapolation. Harmonic enthalpy contributions are included.} 
\label{fig:basis_extrap}
\end{figure}
In Figure \ref{fig:basis_extrap}, we see that convergence of the error for CCSD(T) as a function of the basis set angular momentum is quite slow. At the AVTZ level, errors reach 100 kJ/mol, whereas chemical accuracy is reached for all reactions at the AV5Z level. We also plotted results with the def2-TZVPPD basis set, which performs slightly better than AVTZ in spite of being slightly smaller. Even at the AV5Z level, the agreement of the sum of CCSD(T) and harmonic enthalpy contributions with experimental values is somewhat fortuitous, and the energies are not completely converged yet. Indeed, a cubic extrapolation of the AVQZ and AV5Z correlation energies increases the error from experimental values, while providing results that should be closer to the complete basis set limit. The final extrapolated errors vary between $-2$ and $-10$~kJ/mol.

In conclusion, the combination of harmonic enthalpies and CCSD(T)/AV5Z electronic energies provides combustion enthalpies within chemical accuracy. These accurate results may not transfer to all systems, especially those where high-order correlation effects become important. Provided Ans\"atze on quantum computers can take into account high-order excitations at a sufficiently low polynomial cost, they could provide a better path to chemical accuracy. However, a significant number of qubits would be needed, as we demonstrate in the next section. 

\subsection{Number of qubits}
\label{subsec:nqubits}

In this section we explore how truncation of the orbital active space impacts the combustion energy errors. We take as reference the CCSD(T)/AV5Z electronic combustion energies unless indicated otherwise, and compute the energy difference with respect to the CCSD(T) combustion energy in various basis sets with truncated virtual spaces. In all cases a perturbation theory correction is included to compensate part of the truncation error (see Supplemental Material \cite{Supp_mat}). In our first experiment, we truncate the virtual space by keeping a fixed number of spin-orbitals: 40, 72 or 128.

As was previously reported in the literature \cite{reduced_virt_space_FNO}, canonical virtual orbitals are not an optimal basis for virtual space truncation. Indeed, we observe very large errors in that case even with 128 qubits and the AV5Z basis set, which are the largest active space and basis sets explored, respectively (see Figure %\ref{fig:trunc_canonical} 
S6 \cite{Supp_mat}). The smallest errors range between $-50$ and $-200$~kJ/mol. Moreover, the errors do not converge smoothly as the active space size is increased from 40 to 128 qubits.

\begin{figure}
\center
\includegraphics[width=8.6cm]{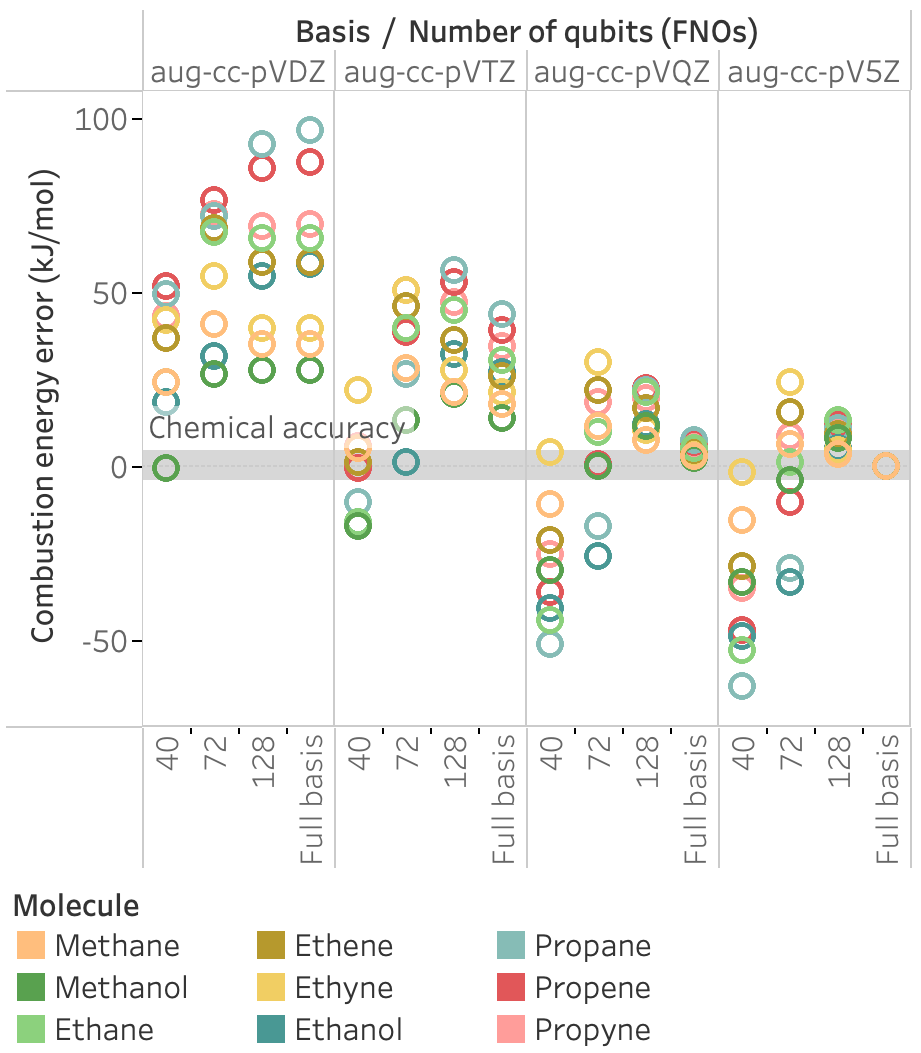}
\caption{ Error relative to CCSD(T)/AV5Z for the FNO method with a fixed number of qubits and frozen core orbitals. Perturbation theory correction is included in the results (see Supplemental Material \cite{Supp_mat}).} 
\label{fig:trunc_FNO}
\end{figure}

A better truncation basis for correlated calculations is provided by Frozen Natural Orbitals (FNOs). In Figure \ref{fig:trunc_FNO}, we plot the errors obtained for different basis sets and active space sizes. We first notice that combustion energy errors visibly converge towards the full basis set limit for each basis examined when going from 40 to 128 qubits. However, even in the largest active space chemical accuracy cannot be reached and the final errors range from 3 to 13~kJ/mol for AV5Z in 128 qubits.

% Note: here and in other figures, the basis sets have their full names instead of the abbreviations I've been using in the text

\begin{figure}
\center
\includegraphics[width=8.6cm]{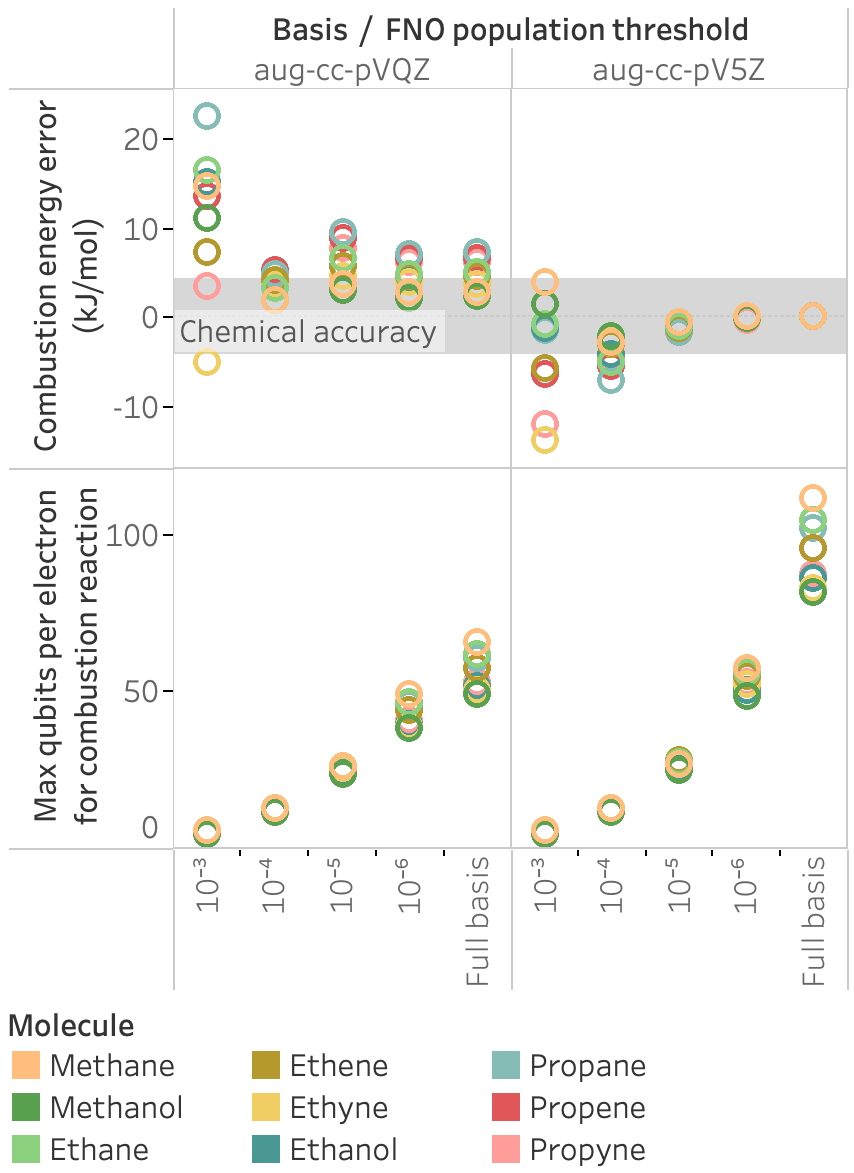}
\caption{ (Top) Error relative to CCSD(T)/AV5Z for the FNO method using the FNO threshold for truncation and frozen core orbitals. (Bottom) Largest number of qubits per active electrons that would be needed to compute the combustion energy for each molecule in the given active space. Perturbation theory correction is included in the results (see Supplemental Material \cite{Supp_mat}).} 
\label{fig:trunc_thresh}
\end{figure}

To facilitate the exploration of active space sizes, our second experiment switches to the FNO occupation threshold as a criterion to select active virtual orbitals. Focusing on AVQZ and AV5Z, we present results for thresholds of 10$^{-3}$, 10$^{-4}$, 10$^{-5}$ and 10$^{-6}$. The upper part of Figure \ref{fig:trunc_thresh} shows that the combustion energy error is indistinguishable from the full basis value at a threshold of 10$^{-6}$. A threshold of 10$^{-5}$ yields a maximum deviation of $-$1.8~kJ/mol from the full basis result, whereas a threshold of 10$^{-4}$ results in a maximum error of $-$7.1~kJ/mol, larger than chemical accuracy.

To connect the FNO threshold to the size of the active space in a transferable way, we plot the maximum number of qubits per active electron for each combustion reaction in the lower part of Figure \ref{fig:trunc_thresh}. This number is obtained by dividing the number of active FNO spin-orbitals by the number of active electrons for each of the target molecule, CO$_2$, H$_2$O and O$_2$ separately. The maximum number of qubits per active electron is the largest result among the four molecules. A threshold of 10$^{-4}$ corresponds to about 13 qubits per electron, which is the number we will use to estimate the size of the active space necessary to reach chemical accuracy. This is an optimistic estimate: the errors we observe are slightly larger than chemical accuracy relative to the full basis limit for AV5Z, but this could be compensated for by including orbital optimization \cite{OO_UCCSD}, or by using some of the qubit reduction techniques \cite{bravyi2017tapering,Steudtner2018} mentioned above.

To conclude, our estimation for the number of qubits $N_{q}$ necessary to obtain accurate dynamical correlation energies is at least 
\begin{equation}
\label{eq:nqubits}
    N_{q} \approx 13 N_{el}
\end{equation}
where $N_{el}$ is the number of active electrons in the system. 

\subsection{Measurement estimation}

In this section, our aim is to estimate the number of measurements needed for a single energy estimation step in the VQE procedure. We consider measurement reduction techniques based on qubit-wise commutativity (QWC) of Pauli terms \cite{min_clique_grouping} and orbital basis rotation \cite{Huggins_tens_fact,OF_patent}, realistic variance estimation, and an efficient orbital basis so that our final estimates reflect conditions close to a large experiment. We aim to obtain empirical extrapolation formulas for the number of measurements for each molecule in our benchmark set. This will allow us to extrapolate the number of necessary measurements for the large qubit active spaces needed for chemical accuracy (see Section \ref{subsec:nqubits}).
We also provide empirical scaling relations for two grouping methods with the size of the system.

\subsubsection{Hamiltonian decomposition methods}

\begin{figure*}
\center
\includegraphics[width=17.5cm]{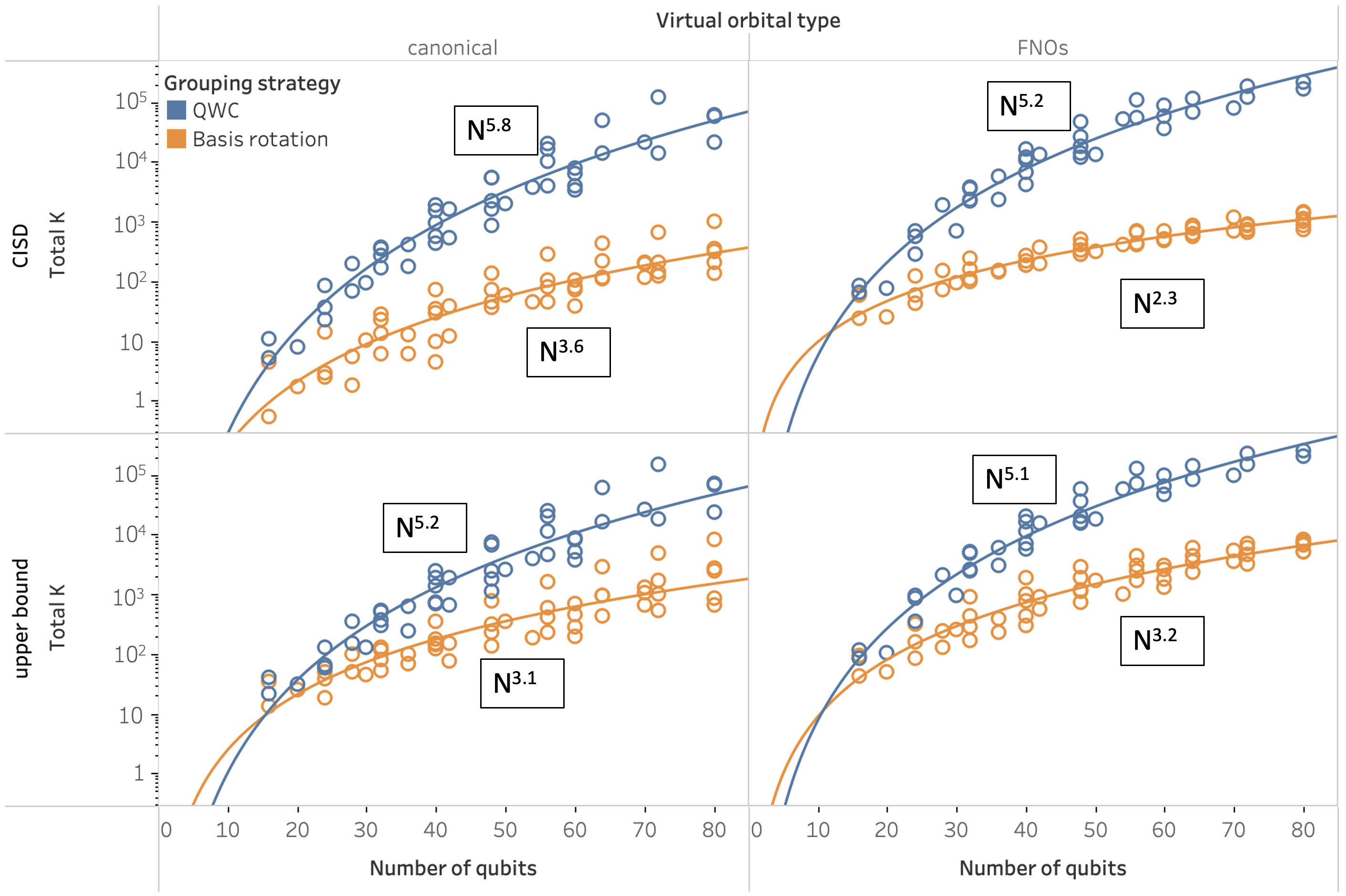}
\caption{Values of K computed for molecules in our benchmark set using QWC grouping (blue) and basis rotation grouping (orange). The top row approximates variances with CISD density matrices and the bottom row sets variances to their upper bounds. Covariances are set to zero in both cases. The left column represents the Hamiltonians in the canonical orbital basis and the right column in the FNO basis. A power law is fit through the data for each grouping method and the obtained exponent is reported next to the curve.} 
\label{fig:average_scalings}
\end{figure*}

We evaluated the Hamiltonian's estimator variance $K$ for QWC and basis rotation grouping with two different orbital bases for the Hamiltonian (canonical orbitals or FNOs) and two different estimates for the variances (upper bounds or CISD, see Section \ref{subsubsec:meas_analysis}), giving a total of 4 different variants for each grouping method. We ran computations for all molecules in our benchmark set (Figure %\ref{fig:molecule_set}
S1 \cite{Supp_mat}), and also included H$_2$O and CO$_2$ that are necessary for computing combustion energies. Due to technical limitations in our code at the time of computation, the open-shell O$_2$ was omitted. For each molecule, we computed different active spaces with an integer number of qubits per active electron up to a total of 80 qubits. This represents the most extensive investigation of the number of measurements in VQE to our knowledge. We fit our results to a power law for each grouping method: 
\begin{equation}
\label{eq:power_fit}
    K = a (N_{q})^b
\end{equation}
where $N_{q}$ is the number of qubits, $a$ and $b$ are fitted parameters. The obtained scaling exponents $b$ are reported next to the corresponding curves on Figure \ref{fig:average_scalings}.

The number of terms in the quantum chemistry Hamiltonian scales as $N^4$, where $N$ is the number of qubits.  However, the QWC grouping method with optimal measurement allocation approximately scales between $N^5$ and $N^6$. The optimal measurement allocation tends to attribute very little to no measurements to terms with very small Hamiltonian coefficients that can safely be neglected. Thus, the observed scaling for QWC grouping only constitutes a modest improvement over the estimated upper bound of $N^6$ for scaling without grouping \cite{meas_scaling_bounds}.

Basis rotation grouping offers better scaling, as hinted by the data presented by Huggins \textit{et al.} for up to 32 qubits \cite{Huggins_tens_fact}. We observe that the scaling varies between $N^{2.3}$ and $N^{3.6}$, a very significant improvement compared to QWC grouping results. We note that the empirically obtained scaling exponents are very close to the optimal bound of $N^2$ derived for the measurement of fermionic 2-particle density matrices \cite{Zhao2020}. In addition, the effect of this improved scaling is already beneficial at low number of qubits, so that QWC grouping never appears advantageous in our computed data. The power law fits indicate that there is a crossover point at which QWC grouping could be preferred, but it only happens below 15 or 20 qubits. Such an example appears in the next section, in Figure \ref{fig:RDMC_res} for 12 qubits. Basis rotation grouping practically always yields a lower number of measurements, however it necessitates the addition of a basis rotation circuit before measurements are performed. Although this circuit has a very shallow depth \cite{kivlichan2018}, in some situations the additional noise induced might become excessive.

To facilitate the comparison of $K$ computed with upper bound and CISD variances, we plotted again the data in Figure \ref{fig:average_scalings} so as to highlight the difference between the two variance estimation methods in Figure %\ref{fig:CISD_vs_upper} 
S7 \cite{Supp_mat}. As expected, this clearly shows that CISD variances always yield a lower number of measurements, albeit by only 20 to 30\% when combined with QWC grouping. With basis rotation grouping, the benefit is significantly larger and reaches a factor 5 to 10. 
This shows that variance approximation is an important aspect to consider when estimating measurements: errors of an order of magnitude can occur when using upper bounds.

The effect of changing the orbital basis of the Hamiltonian from canonical orbitals to FNOs is visualized on Figure %\ref{fig:canonical_vs_FNO}
S8 \cite{Supp_mat}, which contains the same data as Figure \ref{fig:average_scalings} but highlighting the difference of interest in color. When computing the number of qubits needed (see Section \ref{subsec:nqubits}),
we showed that FNOs yield significantly more correlation energy than canonical orbitals for the same number of qubits, which allows chemically accurate results in smaller active spaces. However, this increased accuracy comes at a price since the value of $K$ is systematically higher for FNOs, by a factor of up to 10 in some cases. This is slightly compensated by a lower scaling exponent (compare left and right column of Figure \ref{fig:average_scalings}), that reflects the fact that $K$ saturates faster for FNOs. Indeed, when all virtual orbitals are included, the canonical and FNO spaces are the same and they must have the same $K$.

\begin{table}
\begin{ruledtabular}
\begin{tabular}{ccc}
Molecule & $b$ & $a$ $\cdot 10^{2}$ \\
\colrule
H$_2$O & 1.8 & 45 \\
CO$_2$ & 2.4 & 4.4 \\
Methane & 2.2 & 5.8 \\
Methanol & 2.2 & 9.0 \\
Ethane & 2.5 & 1.9 \\
Ethene & 2.6 & 1.3 \\
Ethyne & 2.5 & 1.6 \\
Ethanol & 2.4 & 3.8 \\
Propane & 2.4 & 2.5 \\
Propene & 2.6 & 1.6 \\
Propyne & 2.7 & 1.0 \\
\end{tabular} 
\end{ruledtabular}
\caption{Fitting coefficients for K for each molecule with $K = a(N_{q})^b$ (Equation (\ref{eq:power_fit})) when using basis rotation grouping, CISD variances and FNOs. $a$ is multiplied by 100 in the table for clarity.}
\label{tab:K_mol_fit}
\end{table}

To obtain extrapolations of the value of $K$ for each molecule, we must choose one of the eight variants investigated. For variance estimation, the CISD approximation is closest to what would be experimentally observed. In spite of the increased number of measurements, we believe the FNO basis is more advantageous since it yields more compact active spaces. Finally, we consider that the circuit fidelity is high enough to afford the orbital rotation circuit from basis rotation grouping. Hence, we fit Equation (\ref{eq:power_fit}) for each molecule using $K$ computed with basis rotation grouping, FNO Hamiltonians and CISD variances. We report our results in Table \ref{tab:K_mol_fit}, and we also plot the fits in Figure %\ref{fig:all_mols_fit} 
S9 \cite{Supp_mat}. The fit to all molecular data presented in Figure \ref{fig:average_scalings} yielded an exponent of 2.3, while for individual molecules $b$ varies between 1.8 and 2.7. Most prefactors $a$ have the same order of magnitude, except for H$_2$O where the prefactor is 5 to 10 times larger than for other molecules. However, H$_2$O also has the lowest exponent. 

The fits presented above represent the scaling of $K$ when increasing the number of qubits for a fixed molecule, which is convenient to extrapolate $K$ to the very large active spaces needed for chemical accuracy. This "virtual scaling" is not the same as the "size scaling", where both the number of active electrons and the number of qubits increase. To investigate size scaling, we fitted Equation (\ref{eq:power_fit}) through our data for increasing numbers of active electrons while keeping the number of qubits per active electrons fixed. We only have enough data to obtain meaningful fits for up to 5 qubits per electrons. Beyond that, the extrapolation gives incoherent results where larger active spaces would need fewer measurements, whereas for 2, 3, 4 and 5 qubits per electrons the obtained scaling is consistent. Overall, size scalings are slightly more favorable than virtual scalings (see Figure %\ref{fig:scalings_qb_per_el} 
S10 \cite{Supp_mat}). The QWC grouping method scales around $N^4$ to $N^{5.5}$ in most cases, whereas the basis rotation method scales between $N^2$ and $N^{2.5}$. Thus, this data suggests that the basis rotation method provides a considerable asymptotic improvement in the number of measurements compared to QWC and related approaches.

\subsubsection{Variance reduction}

\begin{figure}
\center
\includegraphics[width=8.6cm]{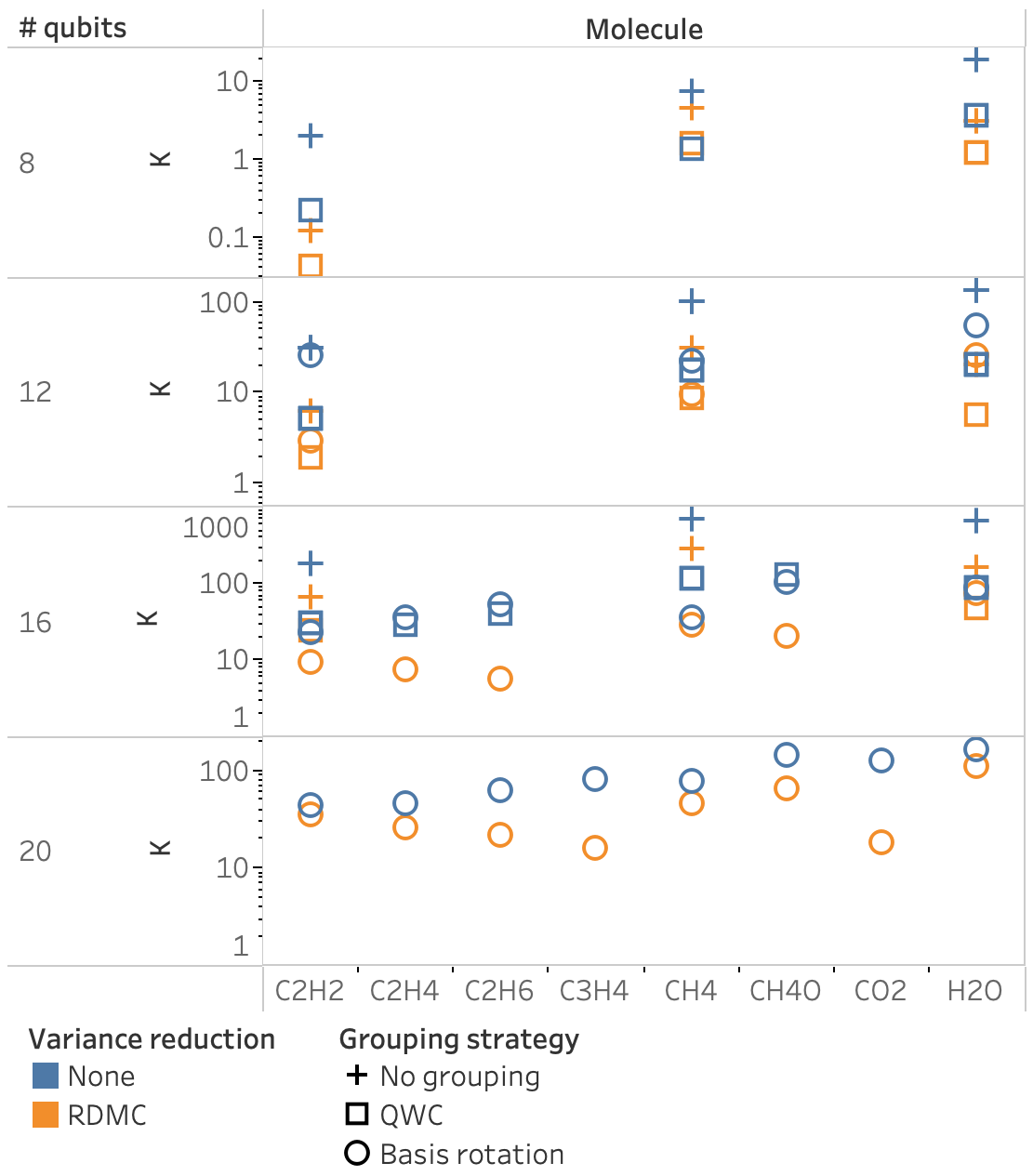}
\caption{Values of $K$ computed for various molecules with no grouping (crosses), QWC grouping (squares) and basis rotation grouping (circles), both with (orange) and without (blue) RDMC. 8 qubit data is freezing 6 electrons for CH$_4$ and H$_2$O, others only freeze core electrons. Hamiltonians were represented with FNOs in the AVTZ basis set and variances estimated from CISD.} 
\label{fig:RDMC_res}
\end{figure}

We now turn to a method that transforms the Hamiltonian to reduce the number of required measurements: the application of fermionic marginal constraints introduced by Rubin \textit{et al.} \cite{marginal_constraints}, that we will abbreviate as RDMC for Reduced Density Matrix Constraints. RDMC as formulated in the original publication scales in principle as N$^{4}$ where N is the number of orbitals. The implementation we are using is based on OpenFermion \cite{OpenFermion} and formulates RDMC as a linear program, which takes significant classical resources. Therefore, we restrict our study to a few molecules and active spaces. We note that we expect an optimized implementation of RDMC to be applicable to much larger systems. Our goal is to obtain an empirical estimation of the improvement in $K$ that RDMC yields. Our results are presented in Figure \ref{fig:RDMC_res}, where we compare the performance of various grouping methods combined with or without RDMC. We see that in all cases, RDMC yields reductions in the values of $K$. The reduction factor obtained is about 3 to 5 when no grouping methods is used. In the case of QWC grouping, the reduction provided by RDMC decreases a bit to a factor of 2 to 3. Basis rotation grouping usually yields the lowest $K$ and has the best scaling with molecular size or number of qubits. Even in this case, RDMC is able to yield an additional improvement to $K$, of approximately a factor of 2. We note that the observed performance of RDMC seems to vary significantly among tested cases, and a factor of 2 is a somewhat conservative estimate. In general, the smaller reduction factors are obtained for larger number of qubits.

For low number of qubits, there are some irregularities in the patterns usually observed. For example, in active spaces of 12 qubits, the QWC grouping method generally performs better than the basis rotation grouping. This also happens for H$_2$O in 16 qubits with RDMC. At these low number of qubits, very few virtual orbitals are included for each active electron, less than one for 12 qubits. This makes it difficult to extrapolate the behavior of the methods examined to large qubit numbers, and highlights the importance of running systematic benchmarks on large enough systems. 

As highlighted in Section \ref{subsubsec:meas_analysis},
our RDMC implementation performs the Hamiltonian transformation in the qubit picture, as was suggested in the original work \cite{marginal_constraints}. In Figure %\ref{fig:RDMC_fermion_qubit}
S11 \cite{Supp_mat}, we compare our results to the original implementation in the fermionic picture, and confirm that the qubit picture implementation systematically yields equivalent or better results. In general, RDMC shows a reduction in measurement count in all cases tested and therefore it could provide practical improvements for the implementation of VQE in the near-term. However, a more extensive analysis of the classical computational cost of this technique and the magnitude of its improvement when scaled to larger systems would improve the current assessment.

\subsection{Overall qubit and runtime requirements}

In this section, we summarize and gather the previous results to obtain estimates for the number of qubits, number of measurements and runtimes required to reach chemically accurate results for the set of investigated combustion reactions. The number of qubits $N_{q}$ is estimated simply from Equation (\ref{eq:nqubits}) and the number of valence electrons in each molecule. The number of measurements is computed as:
\begin{equation}
\label{eq:M}
    M = \frac{K}{2\epsilon^2}
\end{equation}
where $K$ is extrapolated for each molecule separately from Equation (\ref{eq:power_fit}) with $a$ and $b$ taken from Table \ref{tab:K_mol_fit}. The extrapolation takes into account basis rotation grouping, approximated variances from CISD and assumes the Hamiltonian is expressed in the FNO basis. The extra factor of $1/2$ in Equation (\ref{eq:M}) approximately accounts for the additional measurement reduction provided by RDMC on top of the basis rotation grouping. We fix $\epsilon = 0.5$~mHa instead of the usual chemical accuracy of 1.6~mHa. Indeed, we allow 1.1~mHa for additional errors arising from truncation of the active virtual space and from device noise effects. Note that reducing the effect of device noise to below chemical accuracy in general is still a subject of research, and the low error we are assuming can only be achieved on the smallest circuits with the best devices currently.

To convert the number of measurements to actual runtimes, several additional assumptions are necessary. The first and perhaps most speculative regards the Ansatz. Although the UCCSD Ansatz generally yields good results in spite of deficiencies for strong correlation \cite{CC-benchmark}, the corresponding quantum circuit is extremely deep and not appropriate for NISQ devices. Alternatives have been designed \cite{kUpCCGSD,ADAPT_VQE} however we will assume here that we can use a shallower, hardware-efficient Ansatz. Such Ansatz makes use of parametrized entangling gates that are taken to be hardware native or easily compiled to hardware native gates without significant overhead. We are assuming a linear connectivity of the qubit array, in which case a single layer of a hardware-efficient Ansatz is defined as the circuit of depth 2 that entangles every neighboring pair of qubits. We further assume that the number of layers needed to reach the ground state energy scales linearly with the number of qubits, and for the purposes of our estimation, we choose the prefactor in the scaling to be 2. It is likely that this depth constitutes a lower bound for the Ansatz depth that would be necessary in practice. 
Since our extrapolation for $K$ assumes the basis rotation grouping, we also need to add the depth of the circuit for basis rotations, which is $N_{q} - 3$ on a linear array of qubits if $\alpha$ and $\beta$ spins can be transformed independently \cite{kivlichan2018}. 
The final depth of the circuit would then be $5N_{q} - 3$ in terms of two-qubit gates. Our final assumption is that runtime is dominated by execution times of two-qubit gates, which is assumed to be 100 ns, a value on the faster side of current superconducting gate times (see Table 1 in the review by Kjaergaard \textit{et al.} \cite{SC_qubits_review}). The final formula we use to obtain runtimes $t$ in seconds from the values of $M$ and $N_q$ reads:
\begin{equation}
    t = 10^{-7} M (5 N_{q} - 3) 
\end{equation}

\begin{table}
\begin{ruledtabular}
\begin{tabular}{cccccc}
Molecule & N$_{el}$ & N$_q$ & $K \cdot 10^{-3}$ & $M \cdot 10^{-9}$ & $t$ (days) \\
H$_2$O & 8 & 104 & 1.9 & 3.9 & 2.3 \\
CO$_2$ & 16 & 208 & 16 & 32 & 39 \\
Methane & 8 & 104 & 1.6 & 3.2 & 1.9 \\
Methanol & 14 & 182 & 8.4 & 17 & 18 \\
Ethane & 14 & 182 & 8.5 & 17 & 18 \\
Ethene & 12 & 156 & 6.6 & 13 & 12 \\
Ethyne & 10 & 130 & 3.1 & 6.2 & 4.6 \\
Ethanol & 20 & 260 & 24 & 48 & 71 \\
Propane & 20 & 260 & 16 & 31 & 47 \\
Propene & 18 & 234 & 23 & 46 & 62 \\
Propyne & 16 & 208 & 18 & 36 & 44 \\
\end{tabular} 
\end{ruledtabular}
\caption{Estimated runtimes $t$ in days for a single energy evaluation using the number of measurements $M$ from extrapolated values of $K$ (Equation (\ref{eq:power_fit}) and Table \ref{tab:K_mol_fit}), with $\epsilon=0.5$ mHa and the effect of RDM constraints included by a factor of $1/2$ (see Equation \ref{eq:M}). The number of qubits N$_{q}$ is computed from the number of active electrons N$_{el}$ and our empirical estimations of active space size (Equation (\ref{eq:nqubits})). }
\label{tab:runtime_estimates}
\end{table}

We report the results of our runtime estimates in Table \ref{tab:runtime_estimates}. We also plot our estimated runtimes from the computed $K$ values of Figure \ref{fig:average_scalings} and their extrapolation on Figure %\ref{fig:runtimes} 
S13 \cite{Supp_mat}. The picture painted by these runtimes is very pessimistic for VQE. The shortest runtime for energy estimation, for CH$_4$, is 1.9 days. This is in spite of using rather optimistic estimates for the Ansatz depth, the number of qubits needed and neglecting the time for qubit reset, cloud latency times or measurement overheads for error mitigation. Moreover, we highlight again that this is the time necessary for a single energy evaluation. Running the full VQE algorithm involves optimizing the circuit parameters, which requires at least a few dozen to hundreds of iterations even with excellent optimizers. Hence, the total VQE runtime would be about a month for the smallest molecules in our test set. Larger molecules like ethanol already have a runtime of 71 days for a single energy evaluation.

These runtimes originate essentially in the considerable number of measurements necessary to obtain chemically accurate energies for molecules. Even on devices where the error rate would be small enough to warrant reliable VQE execution, the runtime to solution would be prohibitive for molecules in our benchmark set. Parallelization of measurements over several quantum devices is a potential solution, provided all of these quantum devices are sufficiently similar, and the distribution of measurements designed to achieve chemical accuracy. However, parallelization could only bring a constant factor improvement and will not change the scaling of the runtimes with molecular size. In the case of systems dominated by non-dynamical correlation, a smaller active space might be sufficient to demonstrate quantum advantage over classical computing power. A recent paper \cite{industrial_Q_advantage} proposes the chromium dimer with a (24, 24) active space as a potential candidate. At 48 qubits, our extrapolation indicates a runtime of a few hours (see Figure %\ref{fig:runtimes} 
S13 \cite{Supp_mat}), which could allow for a full VQE optimization with considerable effort. However, Hamiltonian coefficients for heavier, strongly correlated atoms like Cr might be larger, which would result in larger values of $K$. Moreover, even if such a computation becomes possible, the transition to practically relevant advantage could require active spaces beyond 100 qubits \cite{industrial_Q_advantage}.

Focusing on the scaling $b$ and omitting the prefactor $a$, our results for the basis rotation grouping technique suggest that VQE has the potential to scale better with system size than methods such as Coupled Cluster. To transform this difference in scaling into an actual practical advantage, research should focus on two directions: 1) developing linear scaling Ans\"atze that provide sufficient accuracy on NISQ devices and 2) improving the measurement techniques, in particular to reduce the dependency of the number of measurements on the required precision. Regarding the first direction, having sufficiently accurate Ans\"atze for VQE with a circuit depth scaling only linearly implies an empirical runtime scaling of $N^{3}$ to $N^4$, which would be advantageous over the scaling of approaches such as CCSD(T). A number of Ans\"atze with linear scaling have been proposed \cite{kivlichan2018,Dallaire2018}, but more studies should be devoted to investigating their representational power for chemical systems of interest, their trainability, and the impact of noise on their accuracy. Along this line, the development and benchmarking of error mitigation techniques is crucial towards achieving sufficient accuracy on NISQ devices. Regarding the second direction, methods that can reduce the dependency of the number of measurements with respect to the required accuracy should be prioritized to make VQE competitive. One such method has been recently proposed by Wang \textit{et al.} \cite{Bulbasaur_1} and Koh \textit{et al.} \cite{ELF} which trade circuit fidelity for a reduction in the number of measurements.

\section{Discussion}
\label{sec:conclusions}

The Variational Quantum Eigensolver (VQE) is a heuristic algorithm, which does not have yet a demonstrated quantum speed-up over classical algorithms for quantum chemistry. Hence, it is of utmost importance to adequately benchmark VQE to evaluate its performance and prospects for quantum advantage. One significant step has recently been made in this direction \cite{industrial_Q_advantage} by identifying what molecules are the most likely candidates for quantum advantage, and in particular for practically relevant quantum advantage.

Here, we outlined a general procedure to assess quantum advantage with a quantum heuristic by carrying out a resource and performance assessment (RPA). We performed a specific RPA for computing a set of combustion energies with VQE, but our general method is also applicable to other variational algorithms. First, it is essential to assess the performance of state-of-the-art classical algorithms to check whether they can solve the problem at hand and estimate the compute resources required. Then, the number of qubits necessary to obtain a solution that is accurate enough should be established. Finally, a rigorous estimation of the number of measurements needed to evaluate expectation values with sufficient accuracy is performed. Measurement requirements are crucial to obtain approximate runtimes, which are ultimately decisive for the practicality of the quantum algorithm.

Our classical benchmarks show that CCSD(T)/AV5Z complemented with harmonic enthalpic corrections is sufficient to reproduce experimental combustion enthalpies to within chemical accuracy. CCSD(T)/AV5Z is taken as our reference energy to estimate the minimal size of the active space that still yields chemical accuracy. Using the well-known FNO method for virtual space truncation, we observe that at least 13 qubits per active electrons must be included to obtain dynamical correlation energies within chemical accuracy. In practice, early quantum advantage will probably be obtained by treating only a small active space on the quantum computer and computing the remaining dynamical correlation energy classically. We believe our results regarding the number of measurements needed are transferable to that case as well.

Our RPA results further show that the number of measurements necessary for QWC grouping scales as $N^5$ to $N^6$, whereas the basis rotation grouping only needs about $N^2$ measurements, at the cost of a small addition to the overall circuit depth. The application of Reduced Density Matrix Constraints on the Hamiltonian in addition to grouping warrants another reduction in the number of measurements by a factor of 2. Unfortunately, the $\epsilon^{-2}$ precision dependence of measurement requirements introduces a very large multiplicative factor. With optimistic assumptions regarding the total circuit depth and the execution time of quantum circuits, estimating a single energy for molecules in our benchmark set to chemical accuracy would take between a few days and a couple of months. Combined with the necessity for a large number of energy evaluations to optimize VQE parameters, this indicates that VQE with sample averaging is not currently practical even for molecules with only a few heavy atoms. 

There are several possible ways to resolve this issue. One is to work on better Hamiltonian decomposition methods, and hopefully achieve reduction in the prefactor or the scaling of the number of measurements needed as a function of the system size. Another would be to work on improving Hamiltonian transformations to reduce the Hamiltonian variance further. Some of these directions have been explored by the authors without significant success. However, a more concrete improvement tackles the $\epsilon^{-2}$ dependence of the number of measurements. Recently, the use of Bayesian techniques combined with Engineered Likelihood Functions \cite{Bulbasaur_1,ELF} offered a way to exploit better device fidelity to reduce the number of measurements, bridging VQE and Quantum Phase Estimation in a practical way. Engineered Likelihood Functions may then be combined with grouping and variance reduction techniques to further reduce measurement requirements and runtime. Any proposed solution to the measurement bottleneck for the application of VQE should be benchmarked on various molecules and active spaces to assess its robustness and scaling with system size.

\begin{acknowledgements}
The authors acknowledge insightful scientific discussions and suggestions from Alex Kunitsa, Peter Johnson, Christopher Brown, and Peter Love, and support from the team of scientists and engineers at Zapata Computing. The authors would like to acknowledge funding from BP for this research project.
J.F.G., M.D.R. and J.R. have stock/options in Zapata Computing, Inc.
% Should we disclose anything else regarding conflicts of interest?
\end{acknowledgements}

J.R., C.B., E.J.D. and C.M.A. conceived the project. J.F.G., M.D.R. and J.R. performed simulations and analyzed results. J.F.G. wrote the initial draft. All authors contributed to the manuscript.

% SI references
\nocite{dynamic_vs_nondyn_corr} 
\nocite{DMRG_for_chemists,DMRG_paper}
\nocite{FCIQMC_dev_appl,QMC_review,QMC_solids}
\nocite{HCI_method,determ_FCI_QMC_alt,CIPSI_benzene}
\nocite{Gidofalvi2005,large_v2RDM_CASSCF}
\nocite{FeMoCo_strong_corr,FeMoCo_DMRG_space}
\nocite{industrial_Q_advantage}
\nocite{CCSD_T_method,cc-pVTZ}
\nocite{df_CCSD}
\nocite{Psi4}
\nocite{MPn-theory}
\nocite{B3,LYP}
\nocite{wB97M_V}
\nocite{aug-cc-pVXZ}
\nocite{bench_NCI_Hobza,noncov_int_benchmark_Lori,S22-set}
\nocite{CCSDT_method}
\nocite{NWChem}
\nocite{general_CC_PT,Err_general_CC_PT,CC_MBPT_TCE}
\nocite{CCSDTQ_method,CCSDTQ_impl}
\nocite{cc_pCVXZ}
\nocite{VDPT2}
\nocite{QFF_approx}
\nocite{GAMESS}
\nocite{def2-basis-sets,def2_diffuse} 
\nocite{CRC_96}
\nocite{MP2_alkane_isodesmic}
\nocite{HEAT_protocol} 
\nocite{MPn-theory}
\nocite{df_CCSD, df-MP2} 
\nocite{FNO_CI_first,reduced_virt_space_FNO,FNO_for_CC}
\nocite{trunc-NO-CCSD-MP2,df_CCSD}
\nocite{marginal_constraints}

%\bibliography{biblio}


\end{document}

\preprint{}
\title{Supplemental Material for "Measurements as a roadblock to near-term practical quantum advantage in chemistry: resource analysis"}

\author{J\'er\^ome F. Gonthier$^{1}$}
\email{jerome@zapatacomputing.com}
\author{Maxwell D. Radin$^1$}
\author{Corneliu Buda$^2$}
\author{Eric J. Doskocil$^2$}
\author{Clena M. Abuan$^3$}
\author{Jhonathan Romero$^1$}
\affiliation{$^{1}$Zapata Computing, Inc., 100 Federal St., Boston, MA 02110, USA}
\affiliation{$^{2}$BP Innovation and Engineering,
150 West Warrenville Road, Naperville, IL 60563, USA}
\affiliation{$^3$BP Innovation and Engineering,
501 Westlake Park Blvd, Houston, TX 77079, USA}

\date{\today}% It is always \today, today,

\maketitle

\section{Classical benchmarks}
\label{sec:Qchem}

In this section of the Supplemental Material, we present technical details of our quantum chemistry benchmark, and additional data and analysis of interest to specialists. Although CCSD(T) is usually considered the gold standard of quantum chemistry, it is important to note that in some classes of molecular systems, for example those exhibiting non-dynamical correlation, methods such as CCSD(T) are not suitable. In these systems, the mean-field solution to the electronic structure problem is qualitatively incorrect. For a more detailed discussion of dynamic versus non-dynamic correlation, we refer the reader to the existing literature \cite{dynamic_vs_nondyn_corr}. Various methods have been developed and are still being perfected to treat systems with non-dynamic correlation with sub-exponential cost on classical computers like Density Matrix Renormalization Group \cite{DMRG_for_chemists,DMRG_paper}, Quantum and Full Configuration Interaction Monte-Carlo \cite{FCIQMC_dev_appl,QMC_review,QMC_solids},
various selected Configuration Interaction schemes \cite{HCI_method,determ_FCI_QMC_alt,CIPSI_benzene}, variational reduced density matrix optimization \cite{Gidofalvi2005,large_v2RDM_CASSCF} etc. However, until now their application has been limited to relatively small active spaces, though much larger than what was previously possible with previous methods. Large reactive metallocenters, of which FeMo-co is a prominent example \cite{FeMoCo_strong_corr,FeMoCo_DMRG_space}, remain out of reach \cite{industrial_Q_advantage}. Thus, in spite of all the progress made in the field of quantum chemistry on classical computers, numerous challenges remain that quantum computers could help solve.

\subsection{Methods}
\label{subsubsec:reac_details}

The geometries of all closed-shell molecules in our benchmark set were optimized at the CCSD(T)/cc-pVQZ \cite{CCSD_T_method,cc-pVTZ} level with the density fitting approximation \cite{df_CCSD} in Psi4 \cite{Psi4}. For O$_2$, the same level of theory was used with a ROHF reference and without the density fitting approximation for technical reasons. Harmonic vibrational frequencies were computed at the same level of theory as the geometry optimization for all molecules.

To assess the performance of some commonly used electronic structure methods, we computed electronic reaction energies with Hartree-Fock, MP2 \cite{MPn-theory}, and the density functionals B3LYP \cite{B3,LYP} and $\omega$B97M-V \cite{wB97M_V} in the very large aug-cc-pV5Z \cite{aug-cc-pVXZ} basis set. For all closed-shell molecules the RHF formalism was used whereas the UHF formalism was employed for O$_2$. The default Psi4 (version 1.2.1) options were used for other aspects of the computations. These calculations correlated all electrons.

In addition to these commonly used methods, we also explored high-level electronic structure corrections in order to converge energies to within chemical accuracy of experimental data. This allowed us to investigate the importance of various high-order contributions. In the following, we abbreviate the cc-pVXZ, aug-cc-pVXZ and aug-cc-pCVXZ basis sets as VXZ, AVXZ and ACVXZ respectively, where X = D, T, Q or 5 for conciseness. For the smallest molecules in our benchmark set, our final electronic energy $E_{final}$ is computed as:
\begin{eqnarray}
\nonumber
    E_{final} &=& E_{CCSD(T)}^{AV5Z} + \Delta E_{CCSDT}^{VTZ} + \Delta E_{{CCSDT(2)}_{Q}}^{VTZ} \\
    & & + \Delta E_{CCSDTQ}^{VDZ} + \Delta E_{core}^{ACVTZ}
\end{eqnarray}
where the subscript identifies the contribution and the superscript the basis set used. $E_{CCSD(T)}^{AV5Z}$ is the density-fitted CCSD(T)/AV5Z electronic energy with core orbitals frozen, except for O$_2$ where the density fitting approximation was not used. In addition, we also computed the CCSD(T) energy in the AVDZ, AVTZ and AVQZ basis sets to verify the convergence of the corresponding reaction energies. The AVQZ and AV5Z results were used to extrapolate the CCSD(T) correlation energy to the basis set limit using a cubic extrapolation formula \cite{bench_NCI_Hobza,noncov_int_benchmark_Lori,S22-set}.

Higher-order effects are estimated through incremental contributions. $\Delta E_{CCSDT}^{VTZ}$ estimates the effect of a full treatment of triples amplitudes \cite{CCSDT_method} relative to CCSD(T) perturbative treatment:
\begin{equation}
\label{eq:corr_T}
    \Delta E_{CCSDT}^{VTZ} = E_{CCSDT}^{VTZ} - E_{CCSD(T)}^{VTZ}
\end{equation}

The additional effect of quadruple excitations is taken into account perturbatively by the $\Delta E_{{CCSDT(2)}_{Q}}^{VTZ}$ correction:
\begin{equation}
\label{eq:corr_2Q}
    \Delta E_{{CCSDT(2)}_{Q}}^{VTZ} = E_{{CCSDT(2)}_{Q}}^{VTZ} - E_{CCSDT}^{VTZ}
\end{equation}
where all energies are computed without the density fitting approximation in NWChem \cite{NWChem}. In the case of propane, the CCSDT(2)$_{Q}$ \cite{general_CC_PT,Err_general_CC_PT,CC_MBPT_TCE} computations proved to be too expensive and the correction is computed with the VDZ basis set instead of VTZ. Note that the corrections (\ref{eq:corr_T}) and (\ref{eq:corr_2Q}) could be computed simultaneously, but we separate them here to gain insight into their individual contributions.

The difference between the perturbative CCSDT(2)$_{Q}$ treatment and the full, iterative CCSDTQ \cite{CCSDTQ_method,CCSDTQ_impl} treatment of the quadruple amplitudes was estimated for all but the largest systems in our benchmark (namely ethane, ethanol, propane and propene) as:
\begin{equation}
    \Delta E_{CCSDTQ}^{VDZ} = E_{CCSDTQ}^{VDZ} - E_{{CCSDT(2)}_{Q}}^{VDZ}
\end{equation}
with all energies computed in NWChem without the density fitting approximation.

Correlation effects for core electrons should be estimated by including core polarization functions in the basis set \cite{cc_pCVXZ}. Our incremental estimation for core correlation effects is computed as:
\begin{equation}
    \Delta E_{core}^{ACVTZ} = E_{CCSD(T),all}^{ACVTZ} - E_{CCSD(T),fc}^{ACVTZ}
\end{equation}
where the subscript "all" indicates that all electrons are correlated whereas "fc" indicates that electrons in core orbitals were frozen. Computations were done with the density fitting approximation in Psi4.
% Note: Psi4 automatically attributed the def2-QZVPP-jkfit basis for the fitting

To obtain reaction enthalpies $\Delta E_{tot}$, it is necessary to add translational, rotational and vibrational enthalpic contributions at finite temperature to $E_{final}$. The total enthalpic correction $\Delta H_{final}$ was computed as:

\begin{eqnarray}
    \Delta H_{final} &=& H_{CCSD(T)}^{harm} + \Delta H_{B3LYP}^{VDPT2} \\
    \nonumber
    &=& H_{CCSD(T)}^{trans} +  H_{CCSD(T)}^{rot} + H_{CCSD(T)}^{vib, harm} +
    \Delta H_{B3LYP}^{VDPT2}
\end{eqnarray}
where $H_{CCSD(T)}^{trans}$, $H_{CCSD(T)}^{rot}$ and $H_{CCSD(T)}^{vib, harm}$ are harmonic translational, rotational and vibrational enthalpies computed at the density-fitted CCSD(T)/cc-pVQZ level and without density fitting for O$_2$. Their sum is the total harmonic enthalpy $H_{CCSD(T)}^{harm}$. $\Delta H_{B3LYP}^{VDPT2}$ corrects the enthalpy for anharmonic vibrational contributions and is computed as:
\begin{equation}
    \Delta H_{B3LYP}^{VDPT2} = H_{B3LYP}^{vib, VDPT2} - H_{B3LYP}^{vib, harm}
\end{equation}
where $H_{B3LYP}^{vib, harm}$ is the harmonic vibrational enthalpy and $H_{B3LYP}^{vib, VDPT2}$ is the anharmonic vibrational enthalpy from VDPT2 \cite{VDPT2} calculations in the Quartic Force Field (QFF) \cite{QFF_approx} approximation as obtained from anharmonic Zero Point Vibrational Energy and frequencies computed by GAMESS \cite{GAMESS}. All these contributions were computed at the B3LYP/def2-TZVPPD \cite{def2-basis-sets,def2_diffuse} level on geometries optimized with the same method, using a Lebedev grid with 99 radial and 590 angular points. All enthalpic contributions are computed at 298.15 K. 

\subsection{Additional data and analysis}
\label{subsec:Qchem_res}

\begin{figure}
\center
\includegraphics[width=8.6cm]{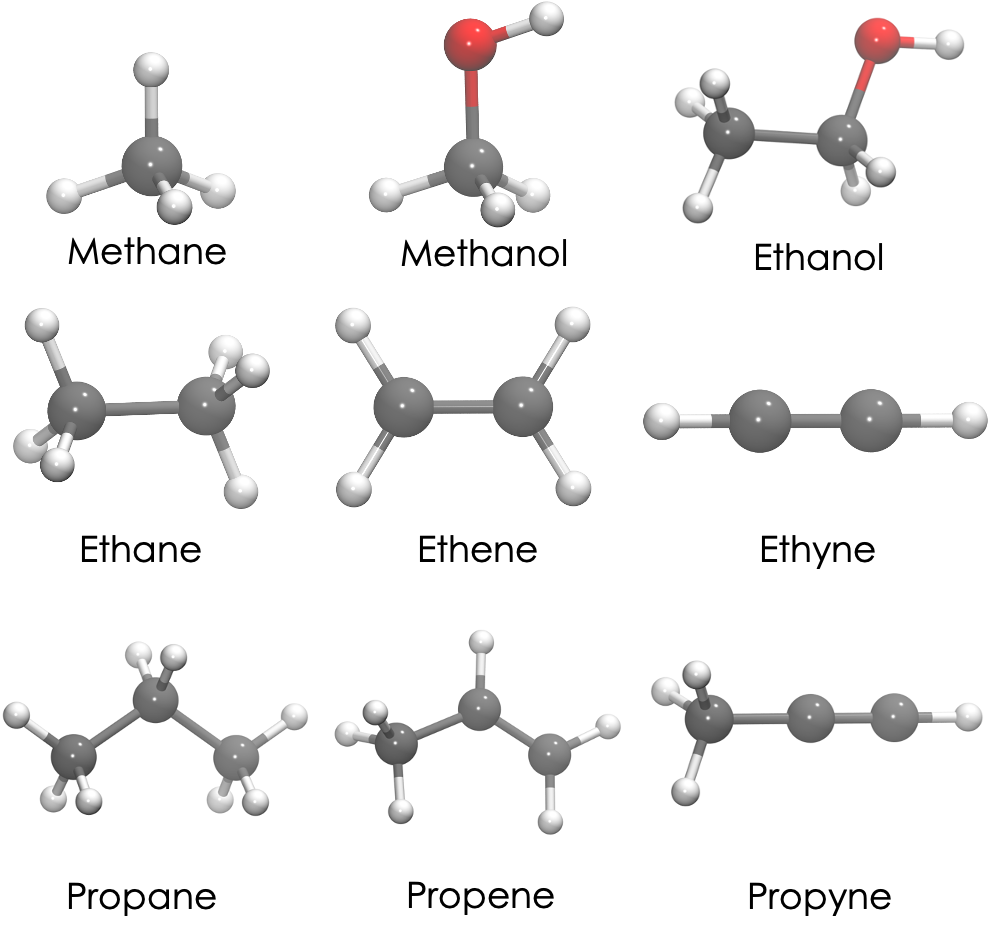}
\caption{Set of hydrocarbons for which we compute combustion enthalpies and corresponding quantum resources.} 
\label{fig:molecule_set}
\end{figure}

In this section, we present additional data relevant to our classical benchmarks and results regarding the performance of quantum chemistry methods for computing the combustion reaction energies for the set of molecules represented on Figure \ref{fig:molecule_set}. The experimental enthalpies of formation that were used to compute the reference combustion enthalpies are presented in Table \ref{tab:experimental_enthalpies}. 

%TODO: should transpose this table
\begin{table*}
\begin{ruledtabular}
\begin{tabular}{cccccccccccc}
Molecule & H$_2$O & CO$_2$ & CH$_4$  & CH$_4$O & C$_2$H$_6$ & C$_2$H$_4$ \\
\colrule
$\Delta H_{f}$ (kJ/mol) & $-$241.8 & $-$393.5 & $-$74.6 & $-$201 & $-$84 & +52.4 \\
\\
Molecule & C$_2$H$_2$ & C$_2$H$_6$O & C$_3$H$_8$ & C$_3$H$_6$ & C$_3$H$_4$ & \\
\colrule
$\Delta H_{f}$ (kJ/mol) &  +227.4 & $-$234.8 & $-$103.8 & +20 & +184.9 & \\
\end{tabular} 
\end{ruledtabular}
\caption{Experimental enthalpies of formation \cite{CRC_96} in the gas phase at 298.15~K and 1~atm for computing combustion enthalpies of molecules in our benchmark set.}
\label{tab:experimental_enthalpies}
\end{table*}

First, we consider the importance of harmonic enthalpic effects and the performance of common, relatively inexpensive quantum chemistry methods.
\begin{table*}
\footnotesize
\begin{ruledtabular}
\begin{tabular}{ccccr}
Molecule & CCSD(T)/AV5Z & $\Delta H_{CCSD(T)}^{harm}$ & CCSD(T)/AV5Z + $\Delta H_{CCSD(T)}^{harm}$ & Experiment \\
\colrule
CH$_4$  & $-809$ & 8.44 & $-800.6$  & $-802.5$  \\
CH$_4$O & $-671.5$ & $-$1.39 &$-672.9$ & $-676.1$ \\
 C$_2$H$_6$ & $-1434.2$ & 7.02 & $-1427.2$ & $-1428.4$ \\
 C$_2$H$_4$ & $-1336.7$ & 13.79 & $-1322.9$ & $-1323.0$ \\
 C$_2$H$_2$ & $-1279.2$ & 21.14 & $-1258.1$ & $-1256.2$ \\
 C$_2$H$_6$O & $-1273.9$ & $-0.43$ & $-1274.3$ & $-1277.6$ \\
 C$_3$H$_8$ & $-2046.7$ & 7.65 & $-2039.0$ & $-2043.9$ \\
 C$_3$H$_6$ & $-1939$ & 14.42 & $-1924.6$ & $-1925.9$ \\
 C$_3$H$_4$ & $-1872.8$ & 20.64 & $-1852.2$ & $-1849$ \\
\end{tabular}
\end{ruledtabular}
\caption{CCSD(T)/AV5Z combustion energies, harmonic enthalpic contributions $\Delta H_{CCSD(T)}^{harm}$to combustion energies, and experimental values \cite{CRC_96} in kJ/mol. All values at 298.15~K and 1~atm, see text for methods.}
\label{tab:combustion_reference}
\end{table*}
The harmonic enthalpy contributions to the combustion energies $\Delta H_{CCSD(T)}^{harm}$ are reported in Table \ref{tab:combustion_reference}. Contributions vary between $-1.39$~kJ/mol for CH$_4$O and +21.14~kJ/mol for C$_2$H$_2$. Since chemical accuracy is generally defined as a maximum error of 4.2~kJ/mol with respect to experimental data, the contribution  of $\Delta H_{CCSD(T)}^{harm}$ cannot be neglected for our set of combustion energies.

\begin{figure}
\center
\includegraphics[width=8.6cm]{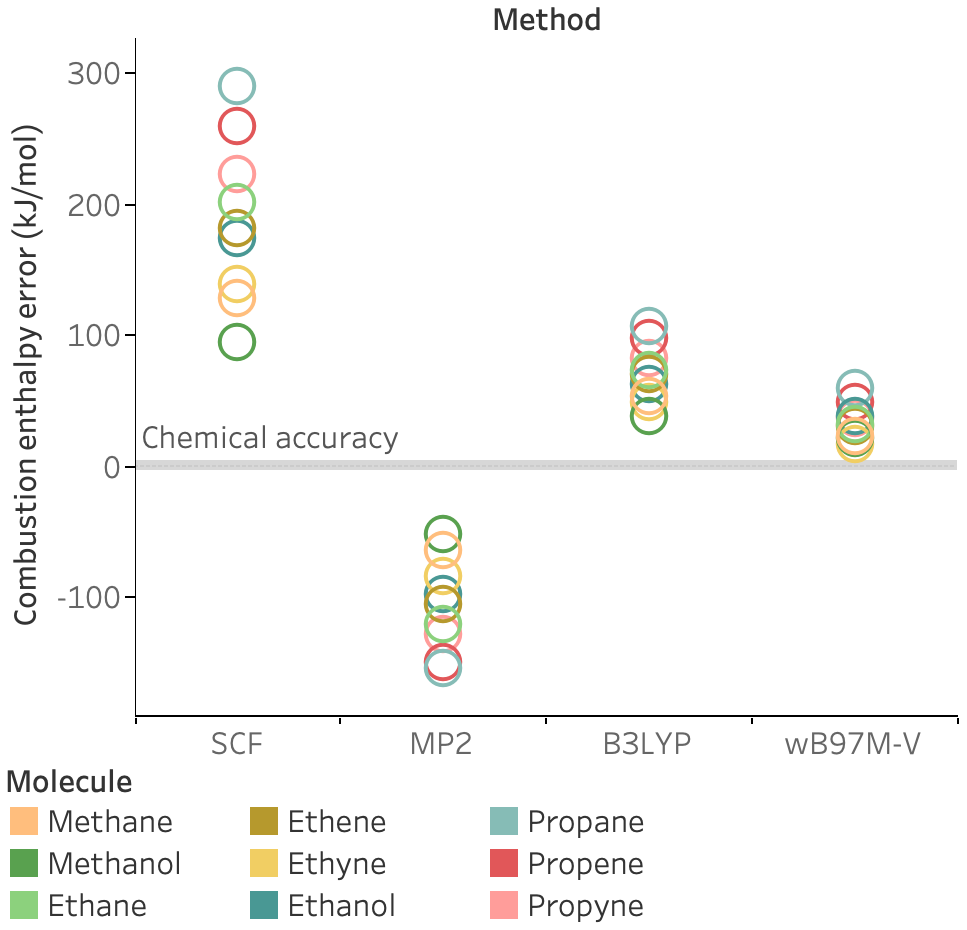}
\caption{Combustion enthalpy errors in kJ/mol using Hartree-Fock, B3LYP, $\omega$B97M-V and MP2 in the AV5Z basis set. $\Delta H_{CCSD(T)}^{harm}$ contributions at the density-fitted CCSD(T)/cc-pVQZ level are included.} 
\label{fig:combustion_methods}
\end{figure}

In Figure \ref{fig:combustion_methods}, electronic combustion energies from Hartree-Fock, B3LYP, $\omega$B97M-V and MP2 are added to $\Delta H_{CCSD(T)}^{harm}$ and the resulting error with respect to experiment is plotted. As expected, the error for Hartree-Fock is the largest and ranges between 100 and 300 kJ/mol because of the neglect of correlation effects. MP2 improves over these numbers, but overshoots the correct value, a behavior which was previously reported \cite{MP2_alkane_isodesmic}.
MP2 errors approximately range between -50 and -150 kJ/mol, well outside the region of chemical accuracy. The popular B3LYP density functional has slightly lower errors of up to 100~kJ/mol. $\omega$B97M-V is a recent functional that provides significant improvements, but still cannot reach chemical accuracy with errors ranging from 17 to 59~kJ/mol. 

\begin{figure}
\center
\includegraphics[width=14cm]{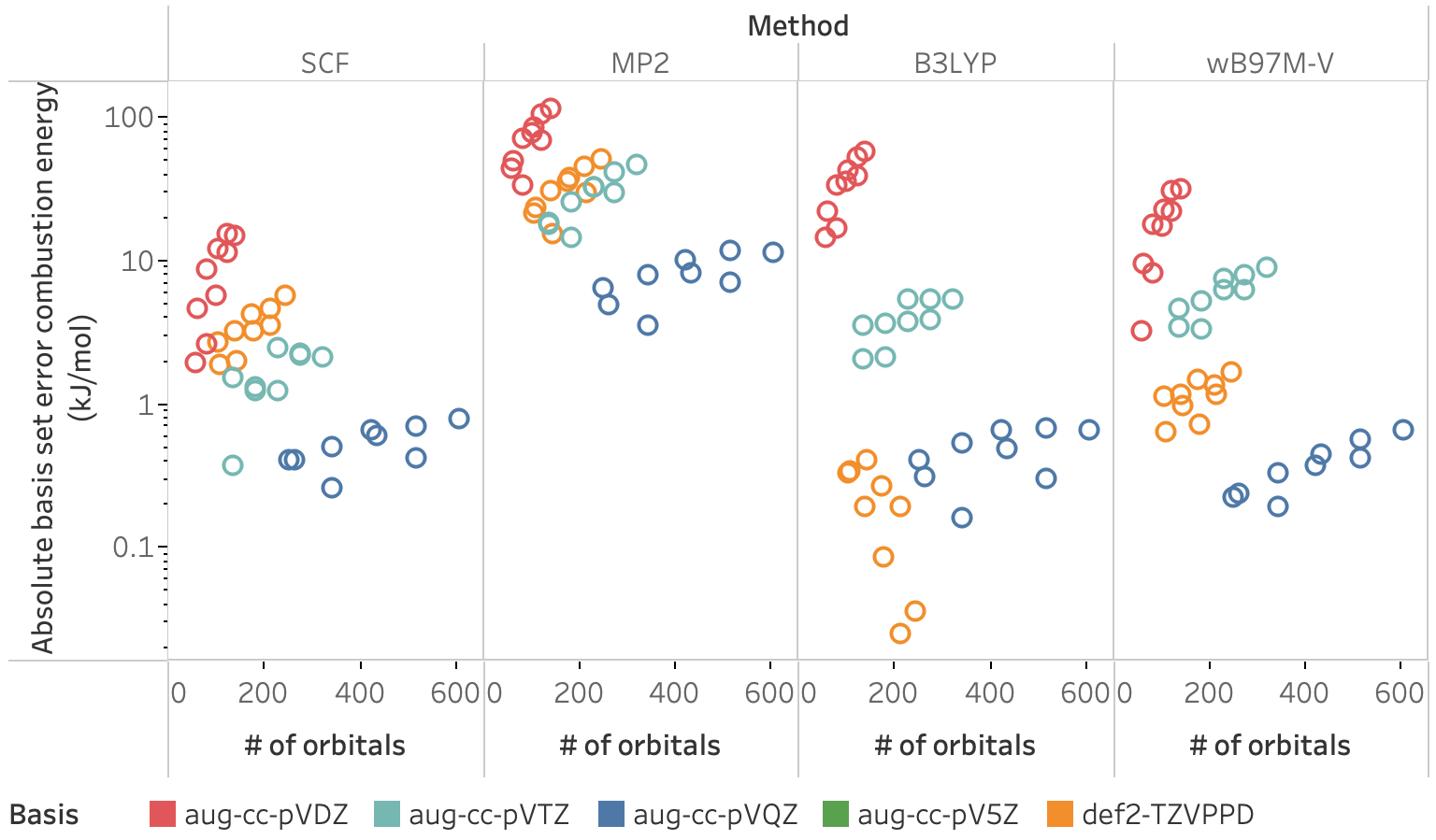}
\caption{Log scale error in combustion energies as a function of the number of orbitals for different methods and basis sets. For each method, the error is computed relative to results in the AV5Z basis set. } 
\label{fig:basis_dep}
\end{figure}

The very large AV5Z basis set used ensures all self-consistent field calculations are converged, as can be seen on Figure \ref{fig:basis_dep} where we plot the basis set error with respect to AV5Z as a function of the number of orbitals for different methods. For most methods, the difference between AVQZ and AV5Z is below 1 kJ/mol. Only for MP2 does this difference reach 10 kJ/mol, which is much smaller than the errors observed on Figure \ref{fig:combustion_methods}. We also observe that the def2-TZVPPD basis set (in orange) is slightly more compact than the AVTZ basis set (in teal), and yields slightly larger errors for SCF and MP2. However, for the DFT functionals B3LYP and $\omega$B97M-V, def2-TZVPPD is both more compact and significantly closer to AV5Z results than AVTZ. 
In conclusion, the performance of these common quantum chemistry methods that can routinely be applied to larger systems is insufficient to reach chemical accuracy.

In the main text, we showed that CCSD(T)/AV5Z with the $\Delta H_{CCSD(T)}^{harm}$ correction reproduces experimental values to within chemical accuracy. However, extrapolating the CCSD(T) correlation energy using the AVQZ and AV5Z bases seems to show that the results are not quite converged with respect to the size of the basis set. In Table \ref{tab:errors_extrap}, we report the results of the AVQZ/AV5Z extrapolation of CCSD(T) and compare them to experimental values. In this case, the extrapolation yields larger errors than using the CCSD(T)/AV5Z numbers (see also Figure %\ref{fig:basis_extrap}
1 in the main text).
%TODO should transpose this table
\begin{table}
\footnotesize
\begin{ruledtabular}
\begin{tabular}{ccccccccccc}
Molecule & CH$_4$  & CH$_4$O & C$_2$H$_6$ & C$_2$H$_4$ & C$_2$H$_2$  \\
\colrule
CCSD(T)/Q5 + $\Delta H_{CCSD(T)}^{harm}$ & -806.1 & $-$678 & $-$1434.1 & $-$1328.1 & $-$1262.4  \\
$\Delta E_{tot, extrap}$ error & $-$2.75 & $-$2.38 & $-$5.84 & $-$4.46 & $-$5.61  \\
\\
Molecule & C$_2$H$_6$O & C$_3$H$_8$ & C$_3$H$_6$ & C$_3$H$_4$ & \\
\colrule
CCSD(T)/Q5 + $\Delta H_{CCSD(T)}^{harm}$ &  $-$1281 & $-$2048.2 & $-$1932.9 & $-$1858.1 & \\
$\Delta E_{tot, extrap}$ error & $-$4.18 & $-$4.65 & $-$6.41 & $-$7.71 & \\
\end{tabular} 
\end{ruledtabular}
\centering
\caption{CCSD(T)/Q5 extrapolated combustion energies combined with harmonic enthalpic contributions $\Delta H_{CCSD(T)}^{harm}$ at 298.15~K and 1~atm. $\Delta E_{tot, extrap}$ includes the sum of high-order, core correlation and anharmonic contributions, similarly to $\Delta E_{tot}$ in the main text, but using the Q5 extrapolated CCSD(T) energies instead of CCSD(T)/AV5Z.}
\label{tab:errors_extrap}
\end{table}
Hence, the extrapolation indicates that converging the CCSD(T) results with respect to the basis set size is likely to yield errors larger than chemical accuracy on combustion enthalpies.

Therefore, we explored the importance of correlation contributions going beyond CCSD(T). These high-order corrections can be computed, albeit at significant cost, by using smaller basis sets. Indeed, the difference between CCSD(T) and high-level methods is expected to converge faster with the size of the basis set than the respective total energies \cite{HEAT_protocol}. We examine these corrections by increasingly high order treatment of amplitudes, and our results are gathered in Table \ref{tab:composite_components}. 

% Table below is a transpose of the next table, keeping it around just in case I change my mind about transposing.
\begin{table}
\begin{ruledtabular}
\begin{tabular}{ccccccc}
Molecule & $\Delta$CCSDT & $\Delta$CCSDT(2)$_Q$ & $\Delta$CCSDTQ & $\Delta E_{{core}}^{ACVTZ}$ & $\Delta$anharmonic & $\Delta E_{tot}$ \\
 (kJ/mol) & (VTZ) & (VTZ) & (VDZ) & (ACVTZ) & & error  \\
 \colrule
CH$_4$ & 0.98 & 2.63 & 0.21 & -3.15 & +0.18 & 2.75\\
CH$_4$O & 0.70 & 1.70 & 0.10 & -2.64 & -0.34 & 2.72\\
C$_2$H$_6$ & 0.99 & 4.39 & N/A & -5.41 & -0.11 & 1.06 \\
C$_2$H$_4$ & 1.49 & 3.86 & 0.31 & -4.86 & -0.16 & 0.74\\
C$_2$H$_2$ & 0.80 & 3.55 & 0.37 & -3.81 & -0.32 & -1.31 \\
C$_2$H$_6$O & 1.23 & 3.49 & N/A & -4.73 & -0.77 & 2.52 \\
C$_3$H$_8$ & 2.00 & 5.32 (VDZ) & N/A & -7.54 & -0.13 & 4.55 \\
C$_3$H$_6$ & 1.92 & 5.74 & N/A & -6.82 & -0.25 & 1.89 \\
C$_3$H$_4$ & 1.24 & 5.39 & 0.46 & -5.60 & -0.10 & -1.81 \\
\end{tabular} 
\end{ruledtabular}
\caption{High-level, core correlations, and anharmonic enthalpy corrections computed in kJ/mol for combustion reactions with the indicated basis set, except for $\Delta$CCSDT(2)$_{Q}$ in the case of propane for which the VDZ basis was used. See Methods for details. $\Delta E_{tot}$ is the sum of all corrections combined with CCSD(T)/AV5Z + $\Delta H_{CCSD(T)}^{harm}$ results, and the error is computed relative to experimental data.}
\label{tab:composite_components}
\end{table}

The $\Delta$CCSDT correction represents the difference between the perturbative treatment of the triples amplitudes in CCSD(T) and the full treatment of these amplitudes which scales as $N^8$. Here, it amounts to at most 2 kJ/mol in the VTZ basis, confirming the accuracy of the CCSD(T) treatment. Next, we introduce perturbatively the effect of quadruples amplitudes through the $\Delta$CCSDT(2)$_Q$ correction that represents the difference between CCSDT and the CCSDT(2)$_{Q}$ perturbative method, scaling as $N^9$. This correction is significant, and contributes up to 5.74 kJ/mol to combustion energies. Omitting the largest molecules, we confirm the accuracy of CCSDT(2)$_{Q}$ by computing the full, $N^{10}$ scaling, CCSDTQ energy at the VDZ level, resulting in a correction of at most 0.46~kJ/mol. 

\begin{figure}
\center
\includegraphics[width=8.5cm]{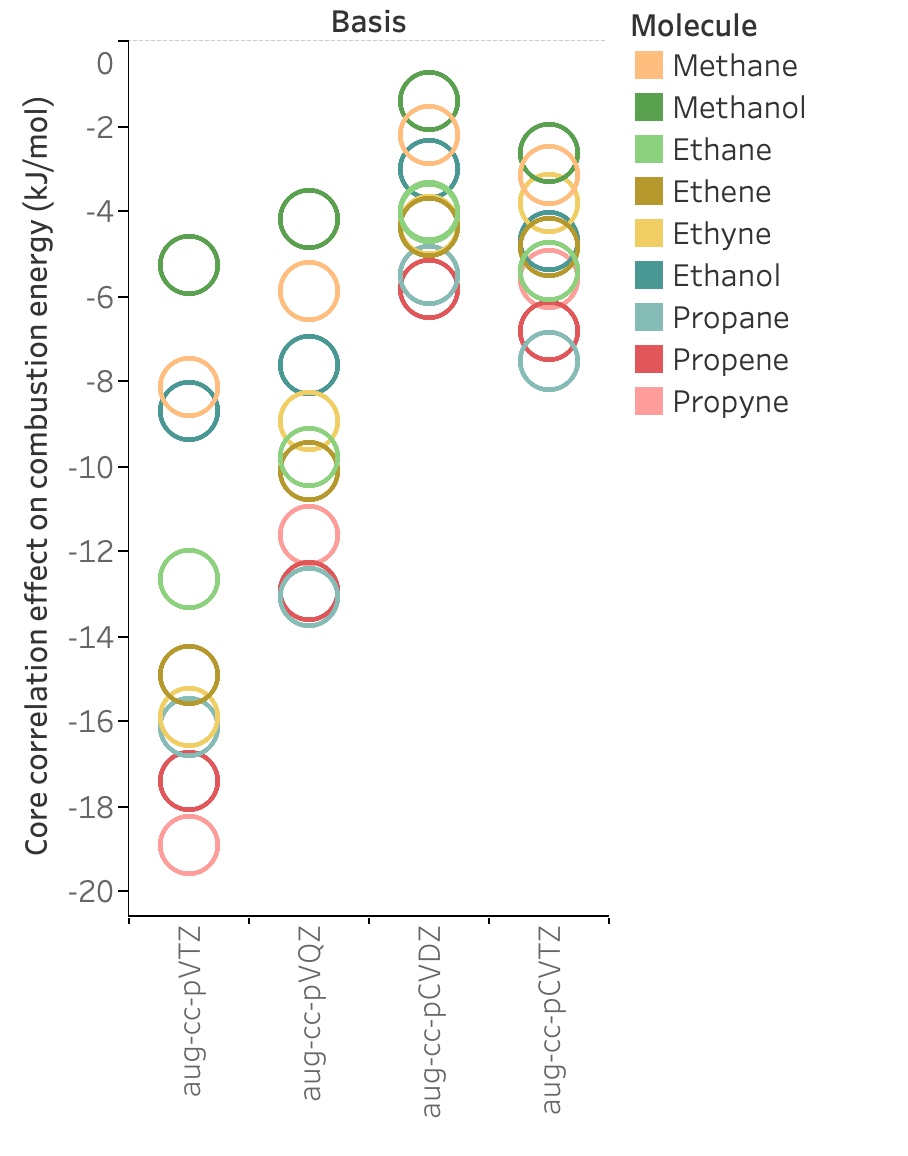}
\caption{Core correlation effects on combustion energies computed with different basis sets. } 
\label{fig:core_correlation}
\end{figure}

The overall effect of high-order excitations is the sum of $\Delta $CCSDT, $\Delta$CCSDT(2)$_{Q}$ and $\Delta$CCSDTQ and ranges between 2.5 and 7.7~kJ/mol. This effect is almost entirely canceled out by the $\Delta E_{core}^{ACVTZ}$ core correlation contributions as observed in Table \ref{tab:composite_components}. In Figure \ref{fig:core_correlation}, we examine the basis set convergence of the core correlation contribution. The ACVDZ and ACVTZ results only differ by 1 or 2~kJ/mol. In spite of using higher angular momentum functions, the results in the AVTZ and AVQZ basis sets that lack core polarization functions vary as much as 6~kJ/mol, and are very different from the ACVTZ results. This highlights the well-known importance of core polarization functions when computing core correlation effects, and confirms the accuracy of the ACVTZ basis set results.

\begin{figure}
\center
\includegraphics[width=8.5cm]{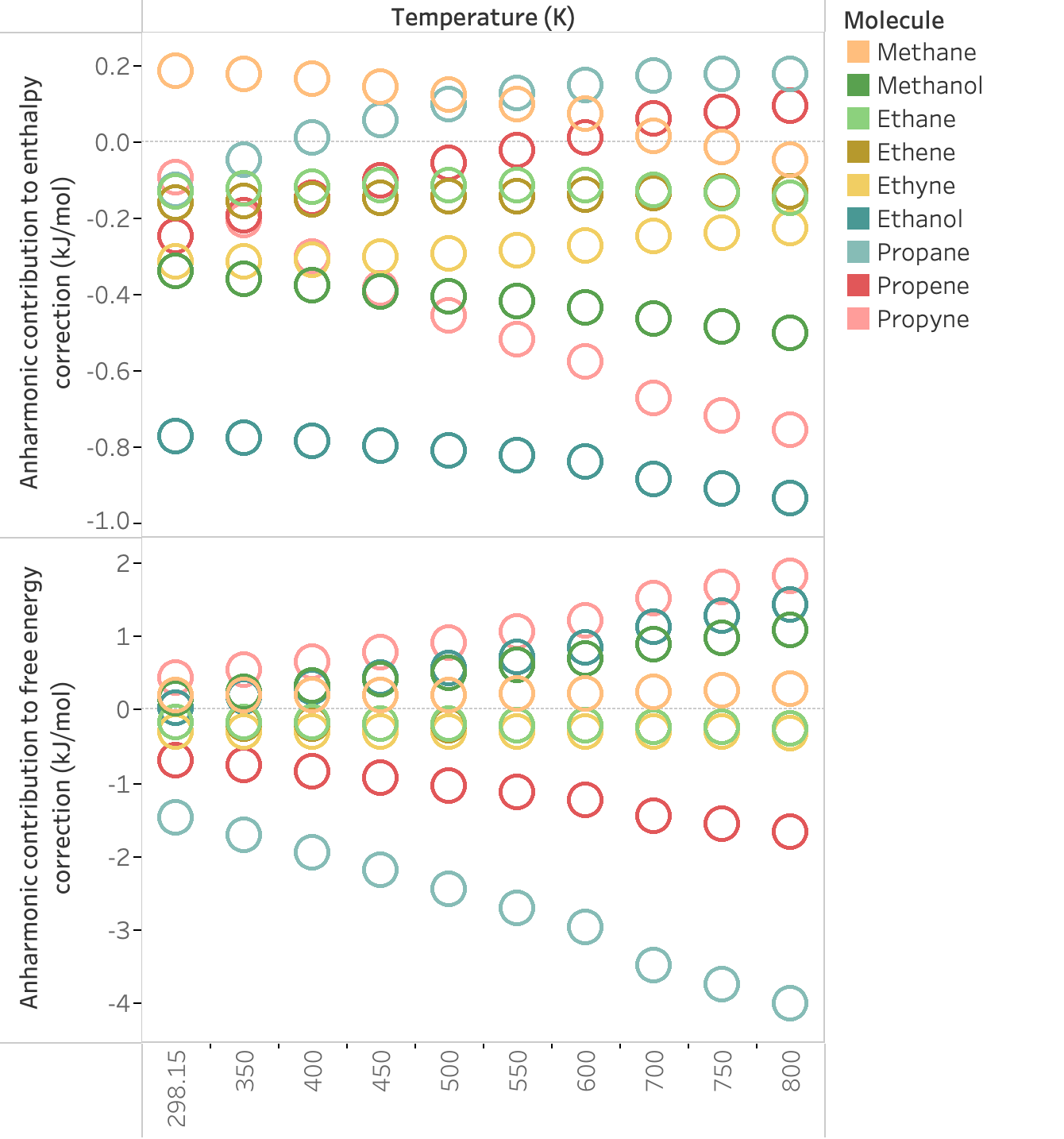}
\caption{ Anharmonic contributions to enthalpy corrections (top) and to free energy correction (bottom) to the combustion energies. These are the purely anharmonic contributions that would be added on top of the harmonic contributions.} 
\label{fig:anharmonic_T}
\end{figure}

Finally, anharmonic contributions to combustion enthalpies are negligible for the molecules considered and range between $-0.77$~kJ/mol and +0.18~kJ/mol. This is expected since vibrational anharmonicity plays a minor role in small rigid molecules, but would become significant in large and flexible systems with soft vibrational modes. In Figure \ref{fig:anharmonic_T}, we plot the purely anharmonic contributions to the enthalpy and free energy of combustion at different temperatures. The temperature dependence of the enthalpy correction is pretty low and is still negligible at 800 K. Because of the temperature-dependent entropic contribution, the free energy correction reaches a value of $-$4~kJ/mol at 800~K, at which point it cannot be neglected for chemical accuracy. Hence, accurate results at high temperatures may also require anharmonic contributions.

Combining all corrections, we obtain an RMSE of 2.4~kJ/mol if the main contributions are computed with CCSD(T)/AV5Z, but of 5.1~kJ/mol if the AVQZ/AV5Z extrapolation of correlation energies is used (see Table \ref{tab:errors_extrap}). By comparison, experimental enthalpies of formation have uncertainties of about 1~kJ/mol or lower for our test set. The lower performance of the extrapolated energies is indicative that some contributions may not be fully converged, or that the AVQZ energies are not reliable enough for extrapolation. In any case, the CCSD(T)/AV5Z result was used in our main investigation.

% JFG: conclusions from original results that relies on high order benchmark results
In conclusion, the accuracy of CCSD(T)/AV5Z stems from a cancellation of errors between high-level correlation effects and core correlation effects. This means that reliably reaching chemical accuracy for large systems is challenging even when only dynamical correlation is present. In cases where the observed error compensation does not occur, even CCSD(T) at the CBS limit might not be sufficient. Provided Ans\"atze on quantum computers can take into account high-order excitations at a sufficiently low polynomial cost, they could provide a better path to chemical accuracy.

\section{Active space selection}

\subsection{Methods}
\label{subsubsec:nqubits_details}

Here, we provide all relevant technical details for our estimation of the minimal number of qubits necessary to preserve the accuracy of the CCSD(T)/AV5Z method. In all cases, we include a correction for the missing correlation energy at the M\o{}ller-Plesset second-order perturbation theory (MP2) level \cite{MPn-theory}, and all computations use the density fitting approximation \cite{df_CCSD, df-MP2}. Hence, the total CCSD(T) energy $E_{CCSD(T)}^{tot}$ is computed as:
\begin{equation}
    E_{CCSD(T)}^{tot} = E_{CCSD(T)}^{trunc} + \Delta E_{MP2}
\end{equation}
where $E_{CCSD(T)}^{trunc}$ is the CCSD(T) energy with virtual orbitals truncated and
\begin{equation}
    \Delta E_{MP2} = E_{MP2}^{full} - E_{MP2}^{trunc}
\end{equation}
where $E_{MP2}^{full}$ is the MP2 energy in the full space while $E_{MP2}^{trunc}$ is the MP2 energy with virtual orbitals truncated. 

Frozen Natural Orbitals \cite{FNO_CI_first,reduced_virt_space_FNO,FNO_for_CC} are obtained by rotating the canonical virtual orbital space to diagonalize the virtual-virtual block of a density matrix including electronic correlation, usually at the MP2 level. The virtual space is then truncated to the desired size by ranking the orbitals by their occupation number, which is usually interpreted as an indication of their importance in the correlation energy. Alternatively, when truncating FNOs by occupation numbers, we used the FNO threshold as implemented in Psi4 for the FNO-CCSD(T) \cite{trunc-NO-CCSD-MP2,df_CCSD} method: virtual orbitals with occupations lower than half the threshold for both the $\alpha$ and $\beta$ spins are excluded from the active space while all others are kept. 

\subsection{Additional data and analysis}

\begin{figure}
\center
\includegraphics[width=14cm]{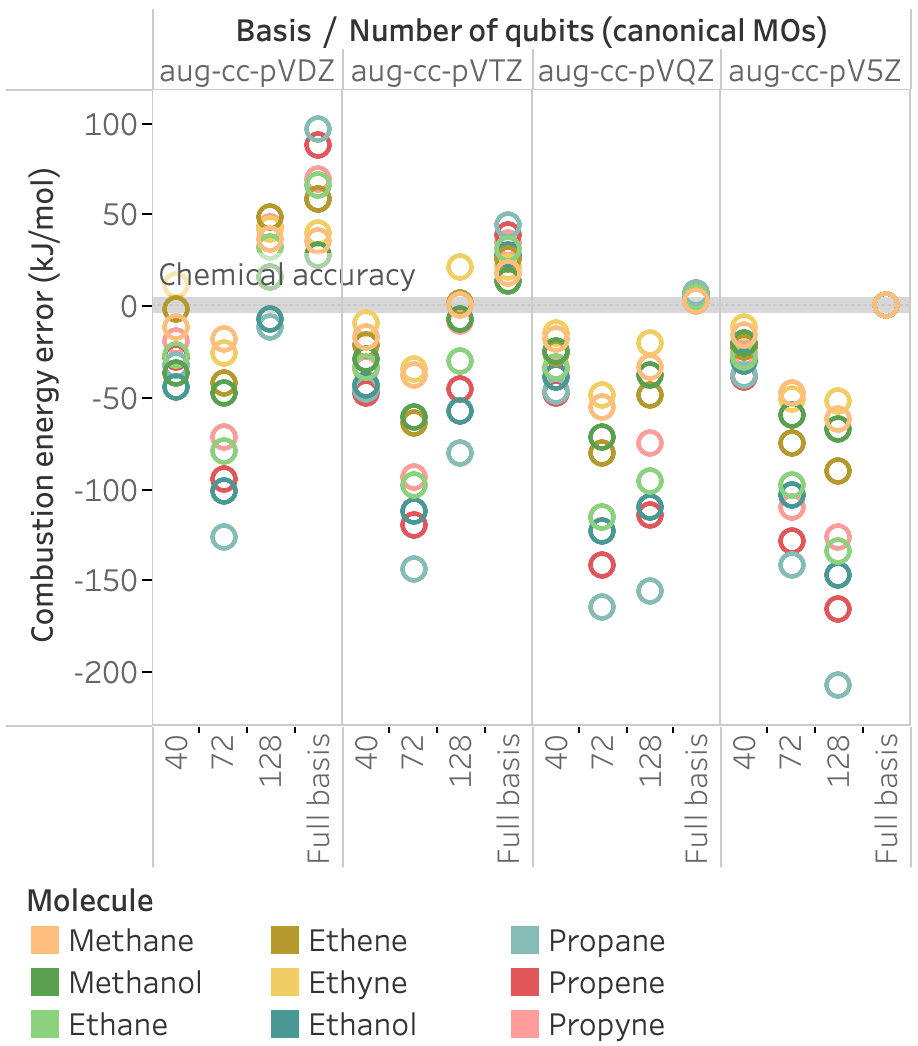}
\caption{For each basis set, we plot the error in combustion energy relative to the CCSD(T)/AV5Z calculation including all orbitals. In each case, we plot the error when keeping 40, 72, 128 or all qubits in the active space, using a canonical orbital basis.} 
\label{fig:trunc_canonical}
\end{figure}

In this section we present additional data regarding our active space selection procedure, which ultimately determines the number of qubits necessary for chemical accuracy. Figure \ref{fig:trunc_canonical} shows that canonical orbitals provide a poor basis for truncation of the active space. Often, the errors do not converge monotonically towards the full basis set limit when increasing the size of the active space from 40 to 128 qubits. Moreover, the 128 qubits results are still significantly different from the full basis result, especially in the larger basis sets.

\section{Measurements}

\subsection{Hamiltonian encoding}

The variance of the energy estimator, which determines the constant $K$, depends on details of the measurement process and the qubit encoding. Here, we consider the widely used Jordan-Wigner encoding of the fermionic Hamiltonian to qubits. The spin-orbitals are ordered by increasing energy or occupation for canonical orbitals and FNOs, respectively. The $\alpha$ and the $\beta$ spins are interleaved, meaning that $\alpha$ orbitals occupy qubits with even indices (starting at 0) and $\beta$ orbitals occupy qubits with odd indices.

\subsection{Additional data for measurement scaling}

\begin{figure}
\center
\includegraphics[width=15cm]{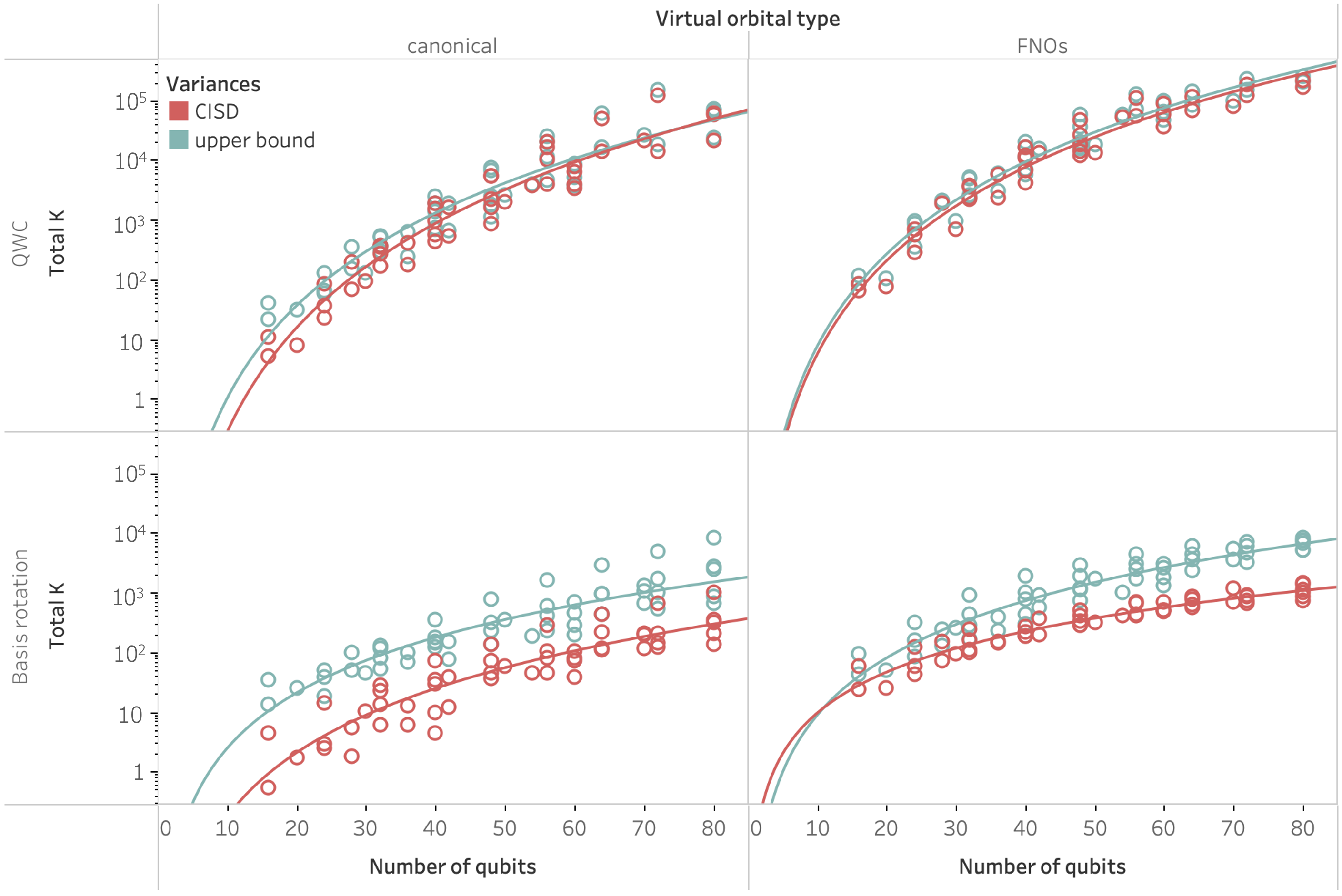}
\caption{Values of $K$ computed for molecules in our benchmark set approximating variances using upper bounds (teal) and CISD data (red). Covariances are set to zero in both cases. The top row uses QWC grouping and the bottom row basis rotation grouping, while the left column represents the Hamiltonians in the canonical orbital basis and the right column in the FNO basis. A power law is fit through the data for each variance approximation method to help visualization.} 
\label{fig:CISD_vs_upper}
\end{figure}

\begin{figure}
\center
\includegraphics[width=15cm]{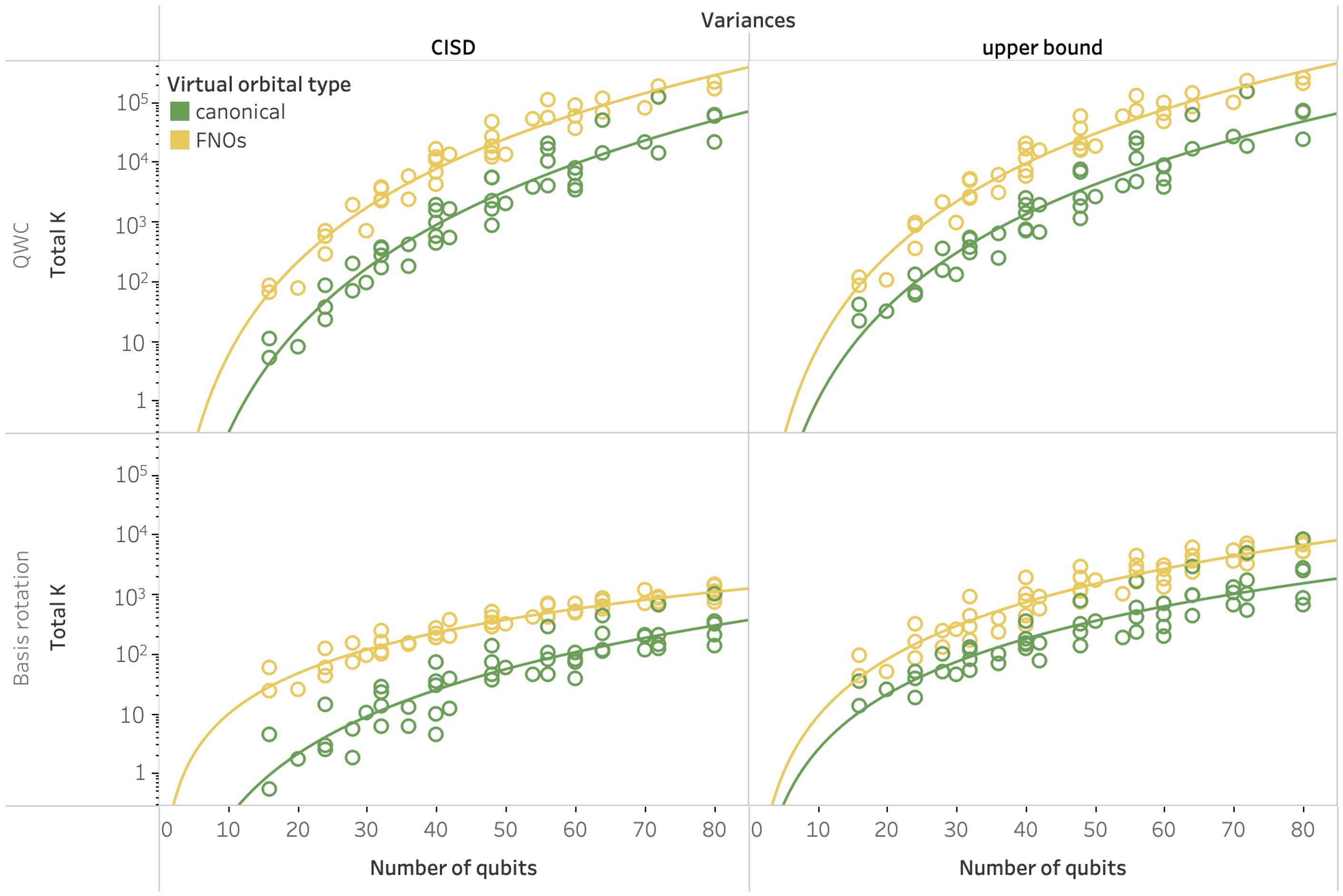}
\caption{Values of $K$ computed for molecules in our benchmark set using canonical orbitals (green) and FNOs (yellow) to represent the Hamiltonian. The top row uses QWC grouping and the bottom row basis rotation grouping. Variances are estimated using CISD in the left column and using their upper bounds in the right column. Covariances are set to zero in both cases.  A power law is fit through the data for each orbital basis to help visualization.} 
\label{fig:canonical_vs_FNO}
\end{figure}

In this section we present some additional plots regarding our measurement scaling results. Figure \ref{fig:CISD_vs_upper} and \ref{fig:canonical_vs_FNO} contain the same data as Figure %\ref{fig:average_scalings}
4 in the main text, but plotted to highlight differences between variance estimation methods and choice of orbital basis, respectively. CISD variances are expected to be closer to experimental variances than upper bounds, and generally yield lower values of $K$, especially in the case of basis rotation grouping. FNOs always yield larger values of $K$ than canonical orbitals, however they are more efficient at recovering correlation energy and allow us to use smaller active spaces and thus less qubits.

\begin{figure}
\center
\includegraphics[width=15cm]{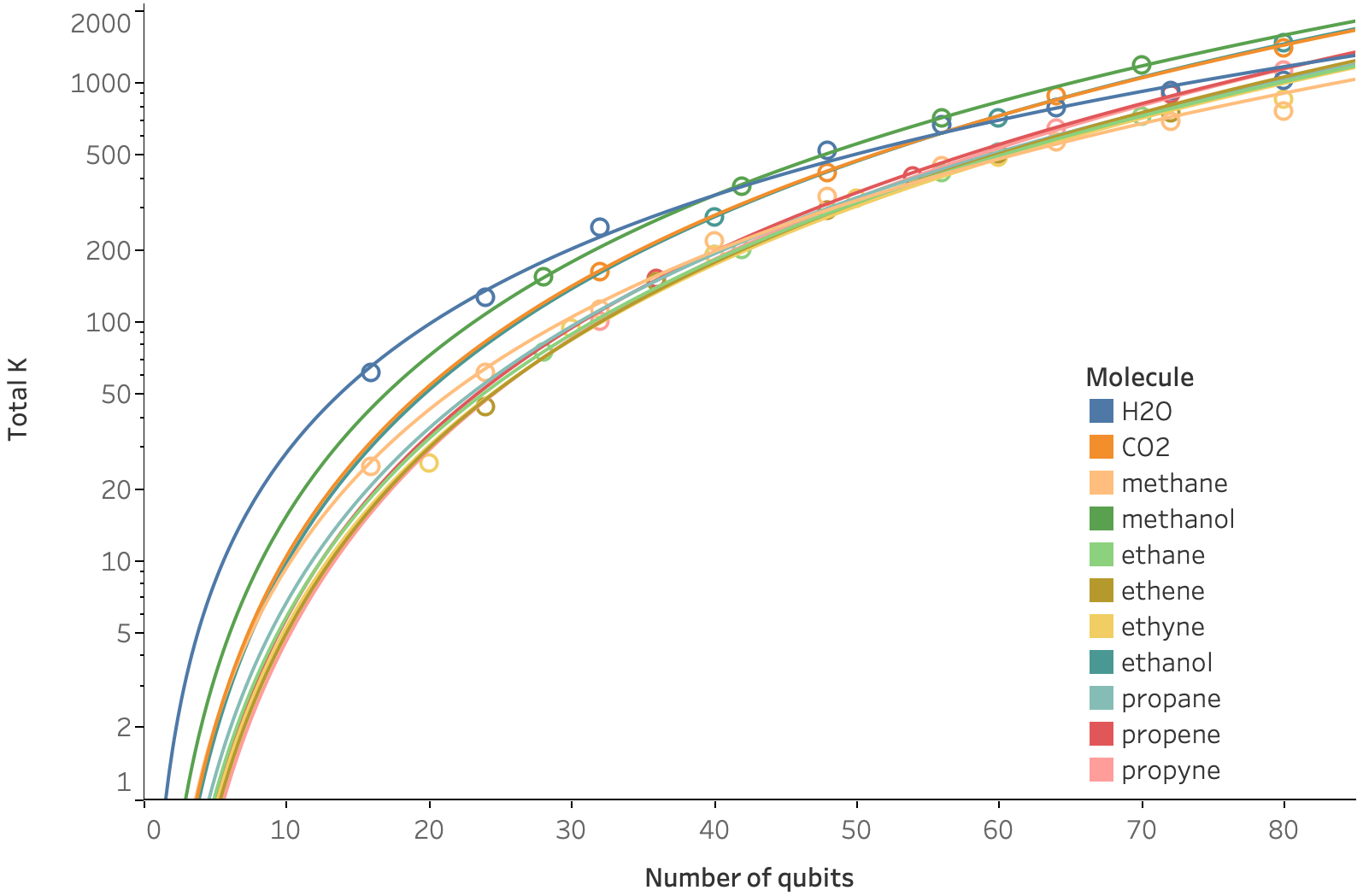}
\caption{Values of $K$ computed for various molecules using basis rotation grouping with CISD variances and FNOs. For each molecule, a power law $K = aN^b$ is fitted through the data. Values of $a$ and $b$ obtained are reported in Table %\ref{tab:K_mol_fit}
I in the main text.} 
\label{fig:all_mols_fit}
\end{figure}

In Figure \ref{fig:all_mols_fit}, we plot the curves obtained by fitting the power law in Equation %(\ref{eq:power_fit})
(8) in the main text to the $K$ values obtained for each individual molecule using basis rotation grouping, CISD variances and FNOs. All curves are similar, which is reflected in the similar values of the coefficients $a$ and $b$ reported in Table %\ref{tab:K_mol_fit}
I in the main text.

\begin{figure}
\center
\includegraphics[width=15cm]{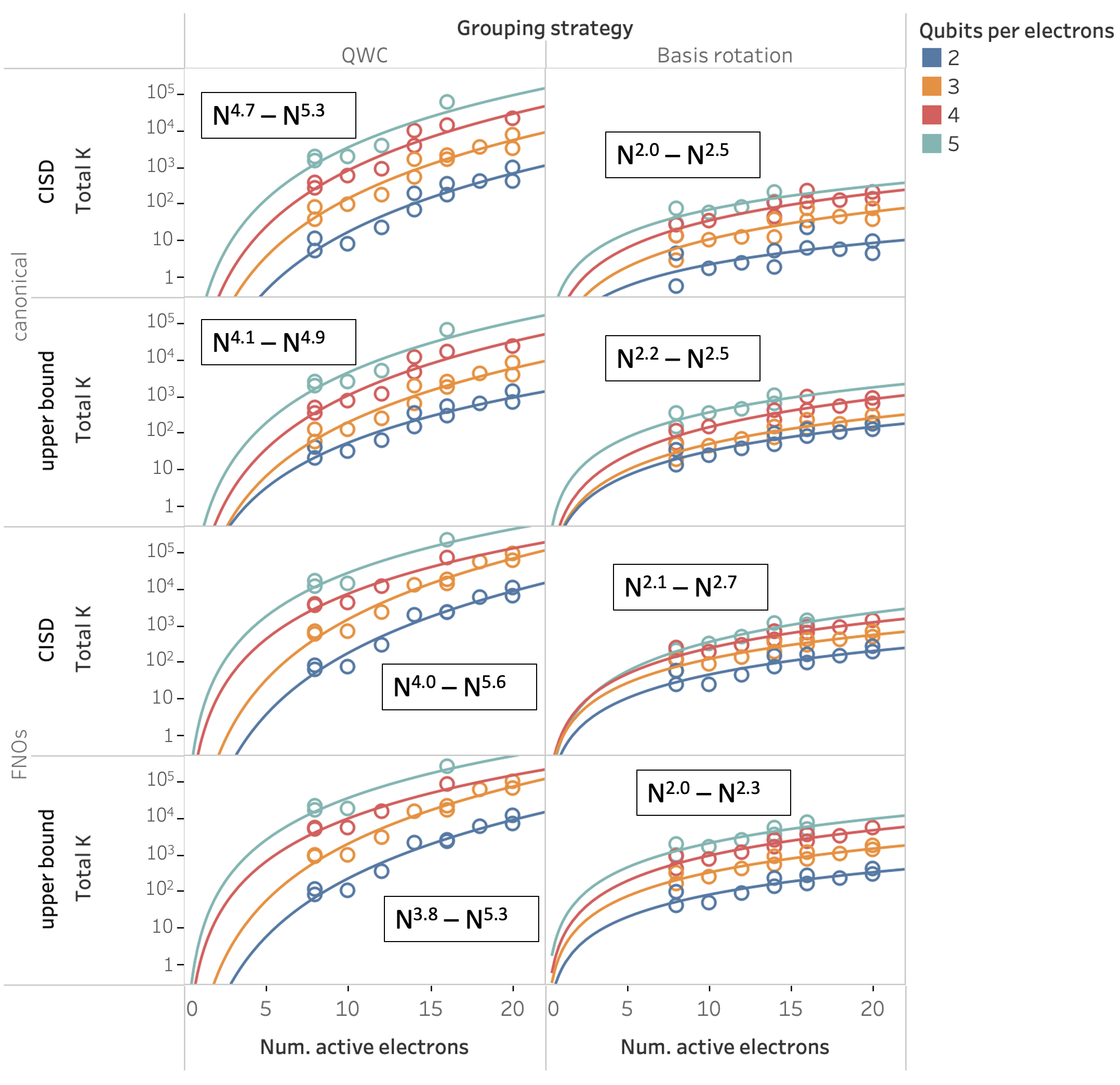}
\caption{Values of K computed for molecules of increasing size, while fixing the number of qubits per electrons to 2 (blue), 3 (orange), 4 (red) and 5 (teal). Left column uses QWC grouping while right column contains results with basis rotation grouping. The top two rows represent the Hamiltonian in the canonical basis and the bottom two in the FNO basis. Finally, variances are estimated by upper bounds or CISD as indicated. A power law is fit through the data for each number of qubits per electrons. The range of exponents obtained for the asymptotic scaling is reported on each plot.} 
\label{fig:scalings_qb_per_el}
\end{figure}

In Figure \ref{fig:scalings_qb_per_el}, we plot the values of $K$ obtained for a fixed number of qubits per electrons, while increasing the number of active electrons. This represents the scaling of $K$ for different variants of grouping methods, variance estimation and orbital bases as a function of the overall size of the system, where both the number of electrons and the number of qubits increase. Beyond 5 qubits per electrons, our data is too sparse for reliable extrapolation and is not represented here. Overall, the scaling coefficients are slightly more favorable than in the case where only the number of qubits is increased and the number of electrons is kept fixed.

\begin{figure}
\center
\includegraphics[width=12cm]{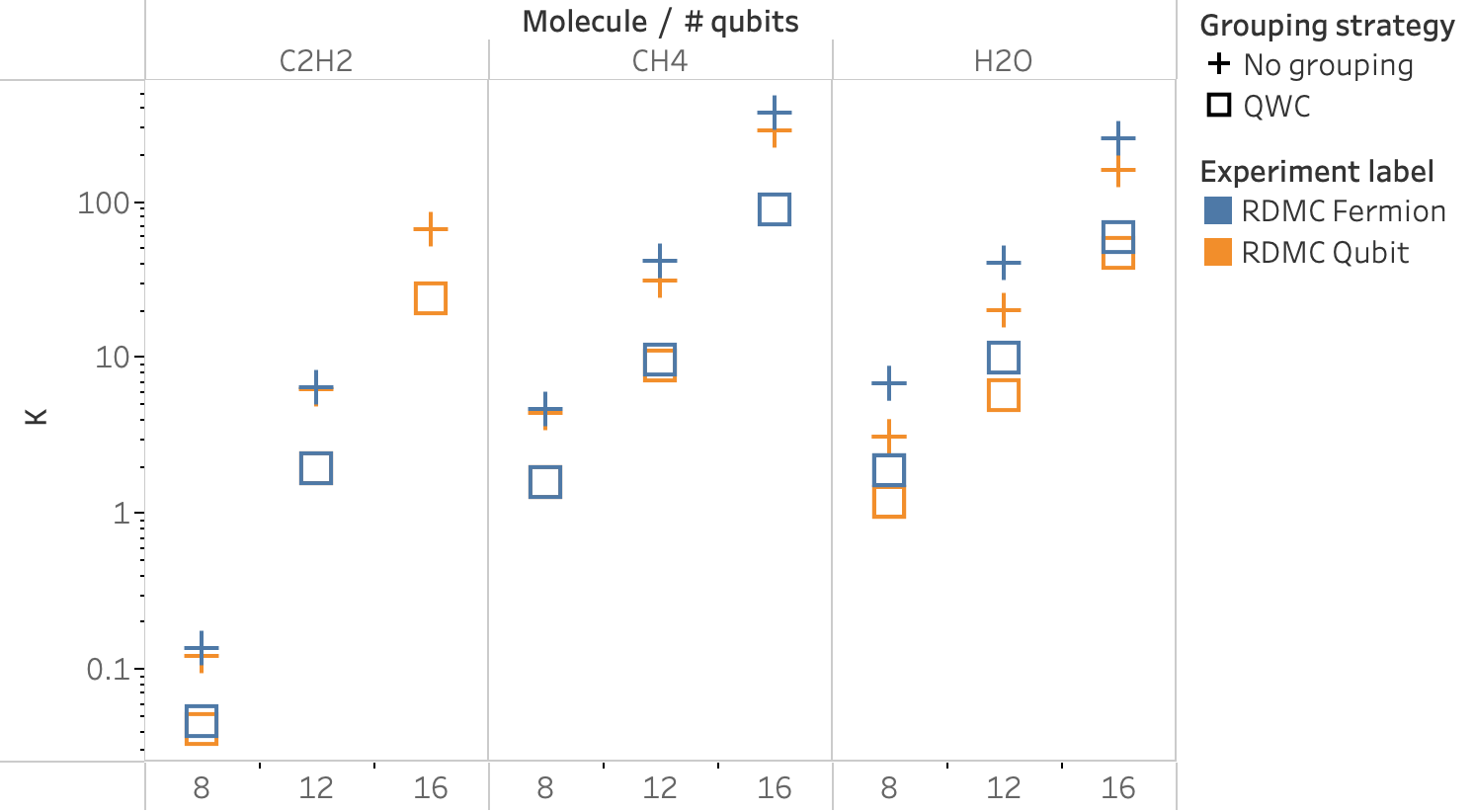}
\caption{Values of K computed for various molecules using no grouping and QWC grouping, combined with RDM constraints in the fermionic picture (blue) or in the qubit picture (orange).} 
\label{fig:RDMC_fermion_qubit}
\end{figure}

Figure \ref{fig:RDMC_fermion_qubit} shows the difference between the implementation of the RDM constraints method in the fermionic and in the qubit picture. The original work \cite{marginal_constraints} employed the fermionic picture implementation, but suggested that the qubit picture might yield better results. Indeed, we observe that our qubit implementation (in orange) performs at least as well or better than the fermionic implementation (in blue). During our experiments, we also observed that the final performance depends on the chosen optimizer.

\begin{figure}
\center
\includegraphics[width=15cm]{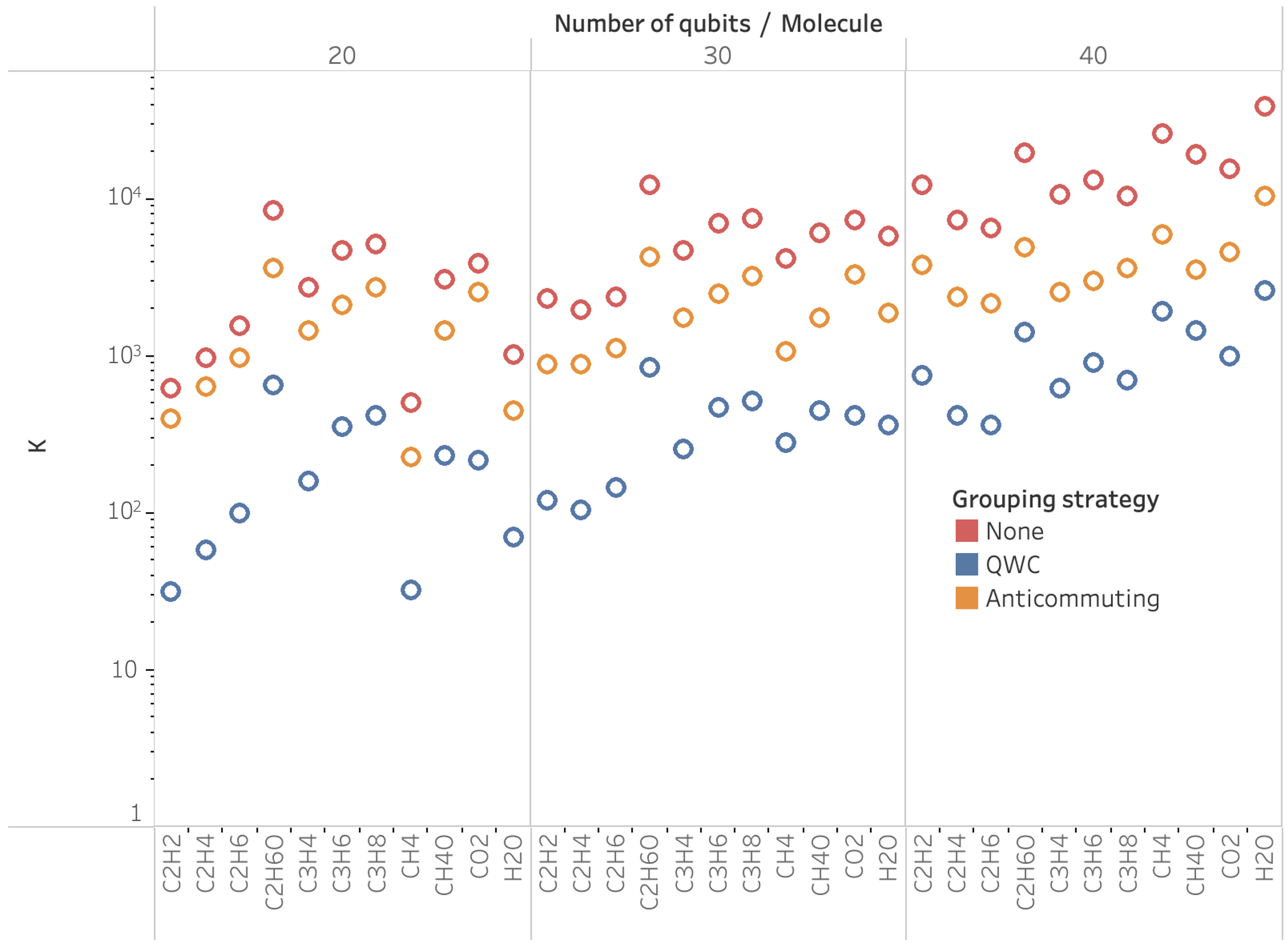}
\caption{Values of K computed for various molecules in active spaces of 20, 30 and 40 qubits, using no grouping (red), QWC grouping (blue) and anticommuting grouping (orange) with canonical orbitals, upper bounds for the variances and zero covariances. Core orbitals were frozen in all cases when computing the Hamiltonians.} 
\label{fig:UD_grouping}
\end{figure}

Figure \ref{fig:UD_grouping} compares values of $K$ computed with no grouping (red), QWC grouping (blue) and anticommuting grouping (orange). For all molecules, the core orbitals were frozen and the Hamiltonian expressed in the canonical orbital basis contained 20, 30 or 40 qubits. The upper bound for the variances was used and covariances were set to zero. The anticommuting groups were determined through the same greedy algorithm than QWC groups, but checking for anticommutativity instead of commutativity. In all cases, anticommuting grouping offers an improvement over no grouping, while being significantly outperformed by QWC grouping.

\section{Additional data for runtimes}

\begin{figure}
\center
\includegraphics[width=15cm]{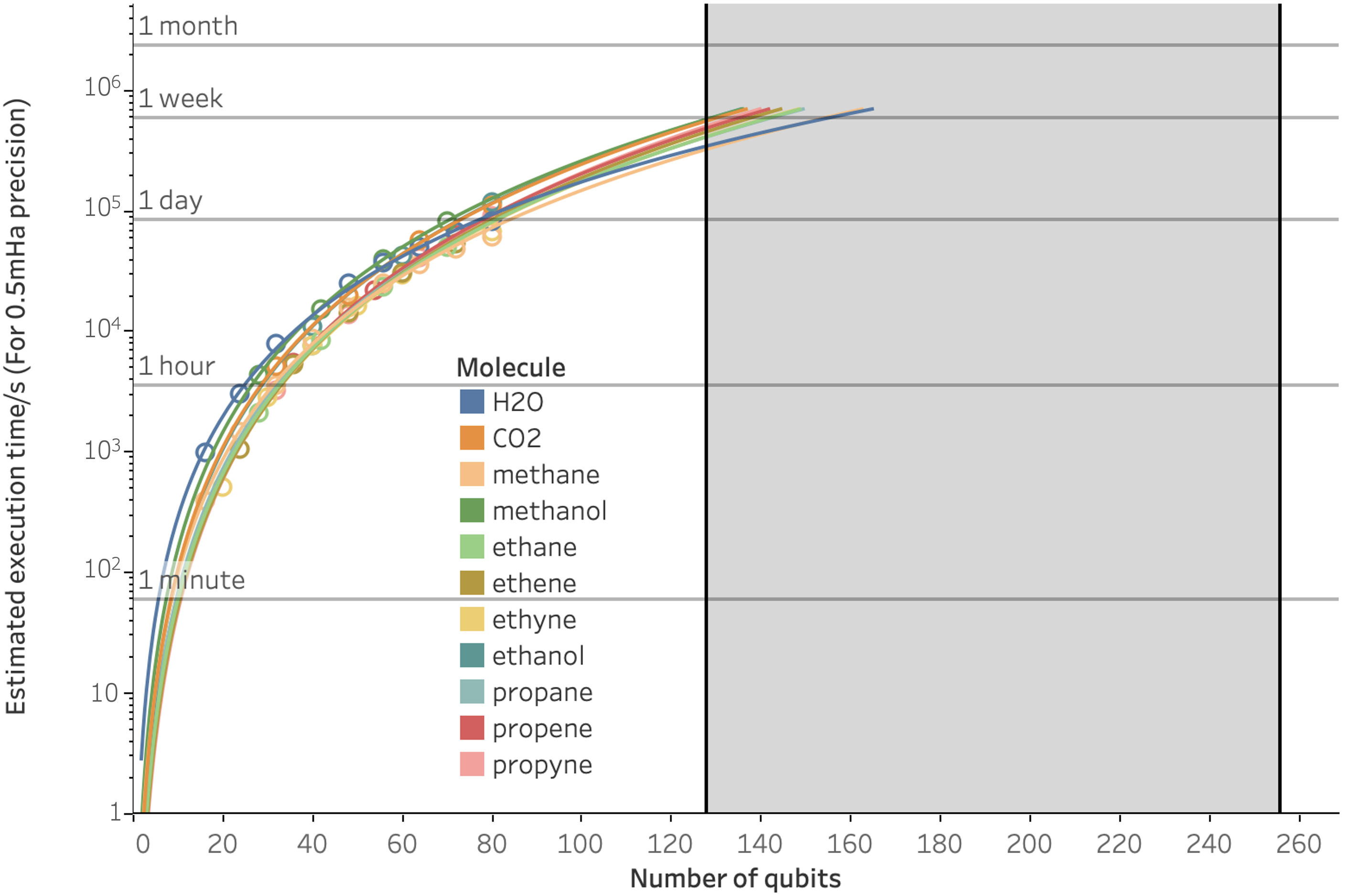}
\caption{Extrapolated runtimes (in s) for molecules in our benchmark set, using the values of $K$ computed for the FNO representation of Hamiltonians, variances from CISD data and basis rotation grouping. An additional factor of 2 improvement is assumed from the RDMC technique, and the number of measurements is computed for a precision of 0.5~mHa to leave room for other uncertainties. Circuit depth is assumed to be $5N_{q} - 3$ two-qubit gates and the two-qubit gate time is assumed to be 100~ns (see main text).} 
\label{fig:runtimes}
\end{figure}

In Figure \ref{fig:runtimes}, we plot our runtime extrapolations for each molecule, using the same estimation method than for Table %\ref{tab:runtime_estimates} 
II in the main text but varying the number of qubits. We observe that at least a few days of measurement times are needed at the lower end of the gray region which indicates the range of number of qubits needed for chemical accuracy in the case of combustion reactions. Assuming a similar scaling, this plot also allows us to estimate a runtime of at least a few hours for the 48 qubit active space corresponding to 24 electrons in 24 orbitals for the chromium dimer. 

